\documentclass[iop]{emulateapj} \usepackage{latexsym} \usepackage{makeidx}
 \usepackage{amsmath} \usepackage{color} \usepackage{upgreek}
\usepackage{morefloats} \usepackage{lineno} \usepackage{setspace}
\usepackage{graphicx}

\shorttitle{Retrieval analysis of 10 Hot Jupiters}
\shortauthors{Barstow et al.} 
\begin{document}

\title{A consistent retrieval analysis of 10 Hot Jupiters observed in
  transmission}
\author{J.~K. Barstow,}
\affil{Physics and Astronomy, University College London, London, UK; Department of Physics, University of Oxford, Oxford, UK}
\author{ S. Aigrain, P.~G.~J. Irwin,}
\affil{Department of Physics, University of Oxford, Oxford, UK} \and
\author{D.~K. Sing} \affil{School of Physics, University of Exeter,
  Exeter, UK} \email{j.eberhardt@ucl.ac.uk}

\label{firstpage}

\begin{abstract}We present a consistent optimal estimation retrieval analysis of ten hot
  Jupiter exoplanets, each with transmission spectral data spanning the visible to
  near-infrared wavelength range. Using the NEMESIS radiative
  transfer and retrieval tool, we calculate a range of possible
  atmospheric states for WASP-6b, WASP-12b, WASP-17b, WASP-19b, WASP-31b,
  WASP-39b, HD 189733b, HD 209458b, HAT-P-1b and HAT-P-12b. We find
  that the spectra of all ten planets are consistent with the
  presence of some atmospheric aerosol; WASP-6b, WASP-12b, WASP-17b, WASP-19b,
  HD 189733b and HAT-P-12b are all fit best by Rayleigh scattering
  aerosols, whereas WASP-31b, WASP-39b and HD 209458b are better
  represented by a grey cloud model. HAT-P-1b has solutions that fall
  into both categories. WASP-6b, HAT-P-12b, HD 189733b and WASP-12b must
  have aerosol extending to low atmospheric pressures (below 0.1
  mbar). In general, planets with equilibrium temperatures between
  1300 and 1700 K are best represented by deeper, grey cloud layers,
  whereas cooler or hotter planets are better fit using high Rayleigh
  scattering aerosol. We find little evidence for the presence of
  molecular absorbers other than H$_2$O. Retrieval methods can
  provide a consistent picture across a range of hot Jupiter
  atmospheres with existing data, and will be a powerful tool for the
  interpretation of \textit{James Webb Space Telescope} observations. \end{abstract}

\keywords{Methods: data analysis -- planets and satellites: atmospheres -- radiative
transfer}

\maketitle

\section{INTRODUCTION} 
Retrieval techniques have been used to great effect for several
decades to invert visible and infrared spectra of solar system planets
and thence infer their atmospheric properties (e.g. \citealt{conrath86,fletcher09}). More recently, these methods have been applied to
observations of transiting extrasolar planets
(e.g. \citealt{lee12,line13,stevenson14,kreidberg14,kreidberg15,benneke15,waldmann15}), although in
many cases there has been considerable degeneracy that has prevented
the determination of a unique solution (e.g. \citealt{barstow13b}). 

Recent observations of transiting hot Jupiters using the Space Telescope Imaging Spectrograph
(STIS) and Wide Field Camera 3 (WFC3) on the \textit{Hubble Space
Telescope} (\textit{HST}) have revealed a variety of atmospheric
characteristics. Most notably, the ten hot Jupiters published in
\citet{sing16} (for additional details of observations, see
\citealt{pont13,line13c,huitson13,mandell13,sing13,sing15,wakeford13,nikolov14,nikolov15,mccullough14})
represent a variety of atmospheres, interpreted as a continuum of clear to cloudy
conditions. We demonstrate in this work that STIS and WFC3
  spectra together cover a sufficient
wavelength range to discriminate between clear atmospheres with
sub-solar water abundances, and atmospheres in which the water feature
is muted by scattering by clouds. 

\citet{sing16} find that the relative transit radii in the visible and
infrared are good discriminators of atmospheric type. Planets with
strong absorption in the visible and weak water vapour abundance
features in the near-infrared are likely to be cloudy, whereas those
with stronger near-infrared water absorption are likely to have clear
atmospheres \citep{sing16,stevenson16,iyer16}. However, it is clear that this is a continuum rather than
a binary state; so, how do the cloud properties of transiting hot Jupiters
vary?

We use an optimal estimation retrieval approach to provide a consistent, data-driven
analysis of the ten hot Jupiter transmission spectra presented by
\citet{sing16}. Whilst optimal estimation does not allow full
  marginalisation over the posterior distribution, it is a fast and
  efficient method that has proven extremely robust for solar system
  studies. Indeed, \citet{line13a} show that, for spectra that are reasonably
  well-sampled in wavelength space, the performance of an optimal
  estimation algorithm is comparable to that of a Differential
  Evolution Monte Carlo method. We choose simple cloud parameterisations to explore
the likely range of cloudy scenarios for each planet. We place
constraints on the cloud top pressure, water vapour abundance and
cloud optical depth of each planet, with varying degrees of confidence
corresponding to the data quality in each case. We compare our
findings with those presented by \citet{sing16} and discuss our
results in the context of cloud formation mechanisms. 

\section{DATA}
All spectral data are taken from \citet{sing16} and references
therein. For all but two of the ten planets, data from
\textit{HST}/STIS, \textit{HST}/WFC3 and warm
\textit{Spitzer}/Infrared Array Camera (IRAC) are combined. No WFC3
observations are currently available for WASP-6b
or WASP-39b, which means that it is not possible to place such
  strong constraints on the atmospheres of these two planets. HD 189733b has additional data from
\textit{HST}/Advanced Camera for Surveys (ACS) and \textit{HST}/Near
Infrared Camera and Multi-Object Spectrometer (NICMOS). The spectral
resolution adopted and number of observations vary from planet to
planet, with obvious implications for the extent to which each
spectrum can be used to constrain atmospheric models. 

All spectral datasets presented by \citet{sing16} were reduced using
consistent systematics models, along with a uniform treatment of limb
darkening and system parameters. A common-mode systematics subtraction
based on the white light curve for each instrument is used, along with
marginalisation of the systematics model following \citet{gibson14}. The
authors estimate that this achieves good reliability in the relative
transit depths between the instruments to within 1-$\sigma$, evidenced
by the good consistency found between three transits depths measured
in the overlapping wavelength regions of the STIS G430L and G750L.  

\section{MODELLING}
For this work, we use the \textit{NEMESIS} radiative transfer and
retrieval code, initially developed for solar system planets
\citep{irwin08} and subsequently used in several analyses of transiting exoplanet
atmospheres \citep{lee12,barstow13b,barstow14,lee14}. \textit{NEMESIS}
uses an optimal estimation algorithm \citep{rodg00} to infer the best fitting
atmospheric state vector from an observed spectrum, and incorporates a
correlated-k \citep{lacis91} radiative transfer model.  

We proceed along similar lines to the retrievals of GJ 1214b spectra
presented in \citet{barstow13b}. In transmission geometry, it is
reasonable to first order to neglect multiple scattering, as the
majority of photons encountering an aerosol particle are likely to be
scattered out of the beam or absorbed given the very long path length
through the atmosphere. Therefore, is it simply the extinction
cross-section of any aerosol that matters, and we do not consider any
effects of the scattering phase function, which vastly simplifies the
parameter space. 

Spectral data included in the models are taken from the sources listed
in Table~\ref{linedata}, and are as used by
\citet{barstow14}. H$_2$-H$_2$ and H$_2$-He collision-induced
absorption is taken from \citet{borysow89,borysowfm89,borysow90,borysow97} and \citet{borysow02}.

\begin{table} \centering \begin{tabular}[c]{|c|c|} \hline Gas & Source\\ \hline
H$_2$O & HITEMP2010 \citep{roth10}\\ CO$_2$ &  CDSD-1000 \citep{tash03}\\ CO &
HITRAN1995 \citep{roth95}\\ CH$_4$ & STDS \citep{weng98}\\ Na & VALD
\citep{heiter08}\\ K & VALD \citep{heiter08}\\ \hline \end{tabular} \caption{Sources
of gas absorption line data.\label{linedata}} \end{table}

Optimal estimation, while a fast and efficient method of spectral
retrieval, provides only a limited exploration of the posterior due to
its dependence on Gaussianity of prior and posterior probability
distributions. In order to explore the parameter space as fully as
possible, a range of retrievals with different cloud conditions and
\textit{a priori} assumptions is performed for each planet. Whilst
this still imposes a restricted parameter space on the problem, we
have attempted to make the exploration as unbiased as possible. The
values of different model parameters adopted are displayed in
Table~\ref{params}, and the rationale for selecting the parameter
space is described further in Section~\ref{radtrans}.

\subsection{Radiative transfer}
\label{radtrans}
Although the datasets from \citet{sing16} have already been analysed
to different degrees, for this work we will make our prior assumptions
about each object as unrestrictive as possible. The \textit{a priori}
model atmospheres for each planet are calculated in the same way,
using only the information available from knowing the planet's period,
mass and radius, and basic properties of the star. The parameter space
explored is presented in Table~\ref{params}.

\begin{table*} 
\centering 
\begin{tabular}[c]{|*{16}{c|}} 
\hline 
Albedo & \multicolumn{4}{|c}{0} & \multicolumn{4}{|c}{0.2} &
                                                           \multicolumn{4}{|c}{0.5}
  & \multicolumn{3}{|c|}{0.8}\\ 
\hline
Gases & \multicolumn{4}{|c}{H$_2$O} & \multicolumn{4}{|c}{+CO$_2$} &
                                                           \multicolumn{4}{|c}{+CO}
  & \multicolumn{3}{|c|}{+CH$_4$}\\
\hline
Cloud (top press. in mbar) & \multicolumn{8}{|c}{Rayleigh} & \multicolumn{7}{|c|}{Grey}
  \\
\hline
Extended & C & 10$^{3}$  & 10$^{2}$ & 10$^{1}$  & 10$^{0}$  &10$^{-1}$  &
                                                                    10$^{-2}$ 
  & U & 10$^{3}$  & 10$^{2}$ & 10$^{1}$  & 10$^{0}$  &10$^{-1}$  &
                                                                    10$^{-2}$ 
  & U\\
Decade & & & 10$^{2}$ & 10$^{1}$  & 10$^{0}$  &10$^{-1}$  &
                                                                    10$^{-2}$&
         & & 10$^{2}$ & 10$^{1}$  & 10$^{0}$  &10$^{-1}$  &
                                                                    10$^{-2}$&
  \\                                      
\hline
Priors & \multicolumn{5}{|c}{Cloud} &  \multicolumn{5}{|c}{H$_2$O} &
                                                                     \multicolumn{5}{|c|}{Na/K}\\
\hline
 & \multicolumn{5}{|c}{0.1$\times$} &  \multicolumn{5}{|c}{0.1$\times$} &
                                                                     \multicolumn{5}{|c|}{0.1$\times$}\\
& \multicolumn{5}{|c}{1$\times$} &  \multicolumn{5}{|c}{1$\times$} &
                                                                     \multicolumn{5}{|c|}{1$\times$}\\
& \multicolumn{5}{|c}{10$\times$} &  \multicolumn{5}{|c}{10$\times$} &
                                                                     \multicolumn{5}{|c|}{10$\times$}\\
\hline
\end{tabular} 
\caption{Parameter space for the 3,600 models used to fit each
  spectrum. Cloud models marked `C' and `U' correspond to clear
  atmosphere and uniform cloud models respectively. All cloud models
  have aerosols distributed with constant specific density in regions
  of the atmosphere where aerosol is present. The total number of
  individual retrieval runs, 3600, comes from 25 cloud models, 4
  temperature profiles, 4 compositions, and 3 tested \textit{a priori}
  values for each of the cloud optical depth, H$_2$O abundance and
  Na/K abundance. 
\label{params}} 
\end{table*}

A key challenge in the interpretation of transmission spectra is the
lack of precise information about the temperature structure, and the high
degeneracy between temperature and baseline pressure at a reference planetary radius in retrievals
(e.g. \citealt{barstow13,barstow13b}). Therefore, we test four
temperature profiles for each planet. Each profile corresponds to a
different value of the Bond albedo. We calculate each planet's equilibrium
temperature using the formula

\begin{equation}
 T_{\mathrm{eq}}=T_{*}(1-a)^{(1/4)}\sqrt{R_{*}/(2D)} 
\end{equation}

where $T_{*}$ is the temperature of the stellar photosphere, $a$ is the
Bond albedo, $R_{*}$ is the stellar radius and $D$ is the orbital
distance. Approximating the atmosphere as a single slab that radiates
equally at upwards and downwards at a temperature $T_{\mathrm{strat}}$,
the temperature of the slab can be calculated by equating the incoming
heat from the star with the outgoing heat from the slab. Assuming an
emissivity of unity, this gives the relation 

\begin{equation}
T_{\mathrm{strat}}=2^{-1/4}T_{\mathrm{eq}}.
\end{equation}

The profile is extended as an adiabat below 0.1 bar, but in
practice we do not expect this region of the atmosphere to be probed
in transmission geometry, so accessible part of the atmosphere is at
temperature $T_{\mathrm{strat}}$.

We also test a variety of simple cloud models. Transmission geometry
is especially sensitive to cloud top pressure and particle
size. Rather than test a series of different sizes of particle with
specific compositions, we test a simple Rayleigh parameterisation and
a simple grey parameterisation. These two extremes correspond to very
small, sub-$\upmu$m sized particles (Rayleigh) and a broad size distribution of large
particles (grey). Within each of these categories, we test 12
different vertical distributions of cloud particles: uniformly
distributed; cloud top at 1000, 100, 10, 1, 0.1 and 0.01 mbar, with
uniform distribution beneath; and cloud top at 100, 10, 1, 0.1
and 0.01 mbar with the cloud base one decade in pressure below. In
each case, where cloud is present it is distributed with a constant
specific density (number of particles per gram of atmosphere) as a
function of pressure. We also test a clear atmosphere model with no
cloud present. 

\begin{figure*}
\centering
\includegraphics[width=0.85\textwidth]{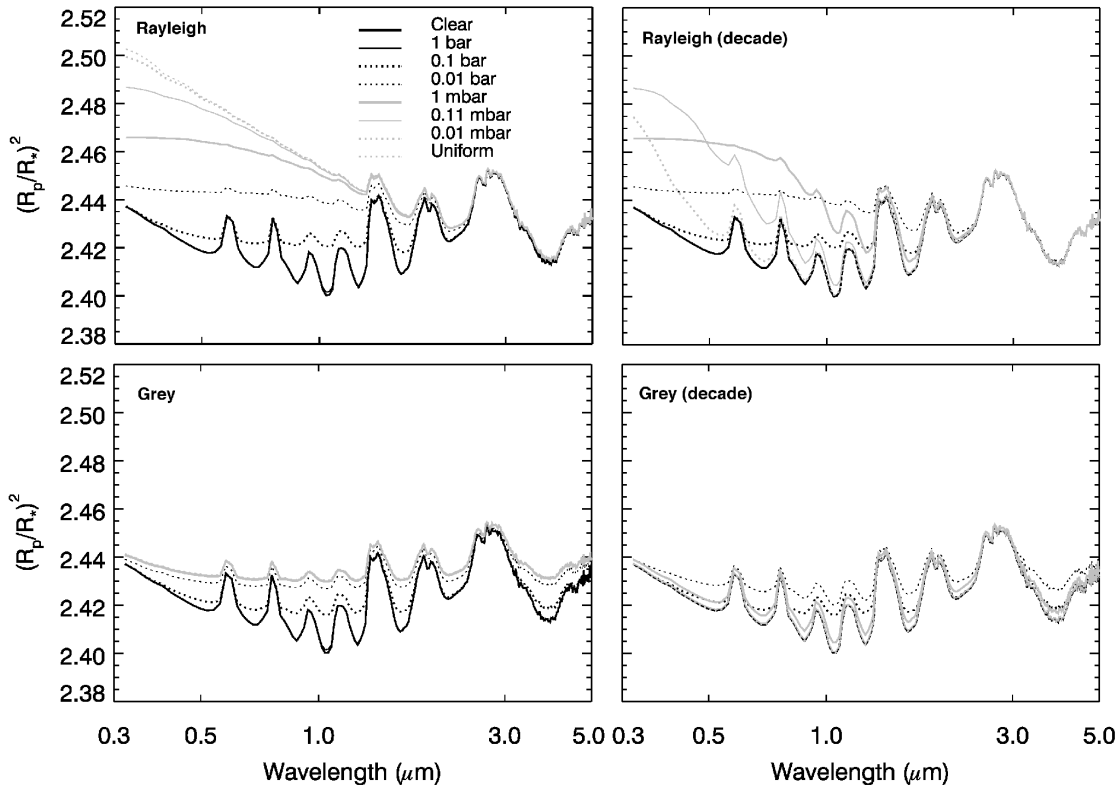}
\caption{The effect of different cloud properties (extinction as a
  function of wavelength,
  cloud top altitude and cloud extent) on transmission spectra
  of hot Jupiters produces a varied range of characteristics. The
  bulk planet properties used are for HD 189733b, and H$_2$O, Na
  and K are the only spectrally active gases included.\label{cloudmods}}
\end{figure*}

The effect of each of the cloud models on the spectrum is illustrated
in Figure~\ref{cloudmods}. It is clear that higher cloud top altitudes
(lower pressures) result in increasingly muted atomic and molecular features in
the visible and infrared, and steeper slopes in the visible. For the cases
where the cloud only spans a decade in pressure, the lower total
optical depth results in a reduced opacity in the red compared with
the blue, making the slopes steeper and the gas absorption
features clearer. Grey cloud produces much flatter spectra than Rayleigh scattering
cloud, but the effect on the infrared molecular absorption features is
similar for both. Rayleigh cloud has a much stronger effect on the
visible spectrum than grey cloud.

\begin{figure}
\centering
\includegraphics[width=0.5\textwidth]{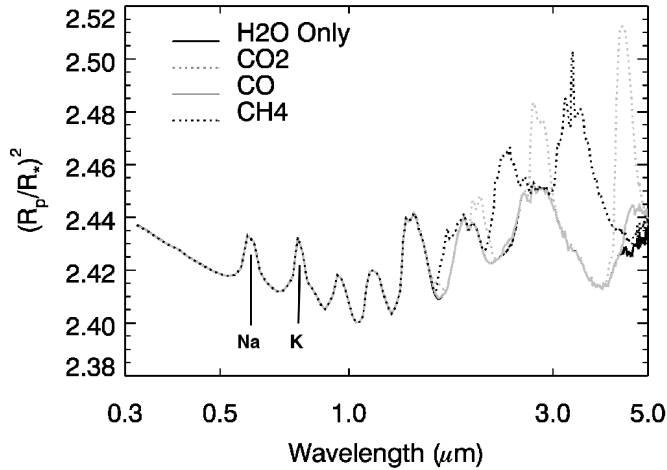}
\caption{The effects of including CO$_2$, CO and CH$_4$ on a hot
  Jupiter transmission spectrum. The H$_2$O-only model (at most
  wavelengths identical to the H$_2$O+CO model) used is the
  same as the clear atmosphere model in Figure~\ref{cloudmods}. The
  addition of extra gases only has measurable effects at the
  wavelengths probed by \textit{Spitzer} (and \textit{HST}/NICMOS for
  HD 189733b). The locations of the Na and K features at optical
  wavelengths are indicated. \label{gases}}
\end{figure}

Due to the limited wavelength coverage of the hot Jupiter spectra
beyond 2 $\upmu$m, there is very little information available about
the presence of molecular species other than H$_2$O. The two broad
band \textit{Spitzer}/IRAC points provide some indication of the
presence or otherwise of absorbers such as CO$_2$, CO and CH$_4$, but
there is insufficient information available in the spectrum to
constrain their abundances. To avoid introducing further degeneracy
into the problem, we test models with H$_2$O only, then models with
H$_2$O plus either CO$_2$, CO or CH$_4$. Each of these gases alters
the relative opacity at the wavelengths probed by \textit{Spitzer} in
slightly different ways (Figure~\ref{gases}).

\section{RESULTS}
We present the results of 3600 retrievals for each of the
ten hot Jupiters in the \citet{sing16} survey. The full results for
each planet are provided as supplementary material; here, we focus
only on the 2\% of tested models that provide the best fits to the
observed spectra. 

We use the reduced $\chi^2$ statistic ($\chi^2_r$) to evaluate the goodness of fit
in each case. This is defined as the $\chi^2$ divided by the number of
degrees of freedom - in this case, the number of spectral points minus
the number of retrieved parameters in each model run. Models
containing an additional molecular absorber to H$_2$O have one less degree
of freedom, so are penalised for additional complexity. We then rank
each model run according to $\chi^2_r$, with lower $\chi^2_r$
values providing the best fits to the measured spectrum. The ranking
is performed by calculating ln($\chi^2_r$) and then normalising,
such that the range for each planet is between 0 and 1. Full retrieved
results against normalised ln($\chi^2_r$) are provided as
supplementary material (Section~\ref{supp}).

\begin{figure*}
\centering
\includegraphics[width=0.85\textwidth]{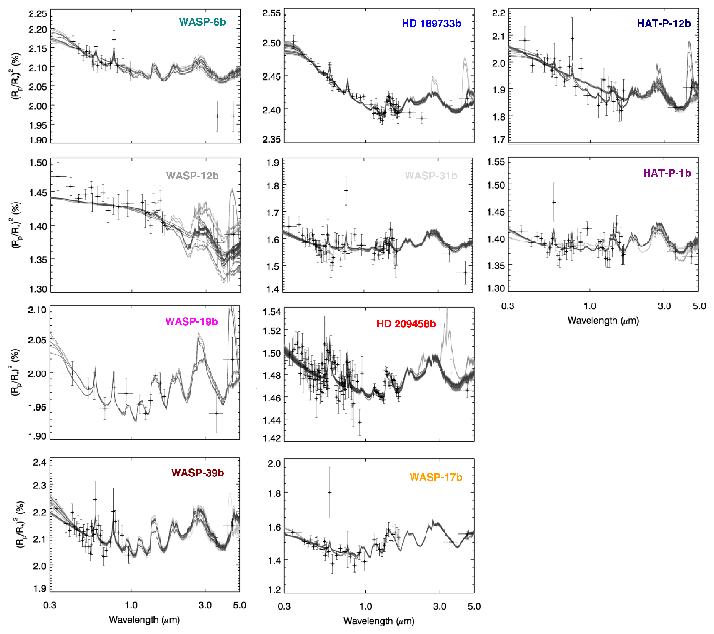}
\caption{Model fits for the best fitting models in each case to
  the observed spectra for each planet. We include only models with a
  normalised ln($\chi^2_r$)$<$0.05. The best fit models are
  shaded almost black, with the models that fit less well presented
  in lighter shades of grey. The width of each spectral channel is
  indicated by a horizontal bar. \label{specfits}}
\end{figure*}

The best model fits to the spectra (normalised ln($\chi^2_r$)$<$0.05) are presented in
Figure~\ref{specfits}. The models are shaded from dark (best) to
light (poorest) fitting based on the reduced $\chi^2$. Spectra with
larger error bars can clearly be fit by a broader range of model
properties, with a lack of \textit{HST}/WFC3 data clearly removing
constraint on the shape of the 1.4 $\upmu$m water band for WASP-6b and
WASP-39b. For some planets, fitted spectra clearly fall into two
distinct categories (e.g. HAT-P-12b, which has one class of solutions
with more opaque cloud and a second with lower opacity in the red), showing that the model parameter
space is bimodal. 

For the most part, spectra are well represented by at least some
models within the family tested. There are however exceptions for
specific parts of some datasets. The \textit{Spitzer} points for
WASP-6b have extremely low transit depths compared with the STIS
measurement; the lack of WFC3 data for this planet makes it difficult
to determine whether this offset is real, or is due to an uncorrected
systematic effect. The spectral fits shown here do not provide a good
match to these data points, but the IRAC
points for WASP-6b are derived from incomplete transits \citep{nikolov14}, so may be
considered less reliable than the STIS data. A better match would be possible for a
model with extremely opaque, high cloud, but these models do not show any Na
or K absorption features as to fit the Spitzer points the cloud needs
to be so opaque and high up that these are obscured completely. An
example of this kind of model can be seen in Figure 1 of
\citet{sing16}. We chose the set of solutions presented here on the
assumption that the detection of Na and K is more robust than the IRAC data points. 

Another clear discrepancy can be seen in the WASP-31b spectrum, where
the 4.5 $\upmu$m IRAC data point cannot be fit by any models in the
family. It is difficult to find a scenario in which the 4.5 $\upmu$m
IRAC point has a transit depth so much smaller than both the 3.6 $\upmu$m
point and the WFC3 data, and the same issue can be seen in Figure 1 of
\citet{sing16}. In general, the error bars on the IRAC points are
larger than those on the STIS points, but the individual points carry
increased weight as they provide the only information available at
wavelengths longer than 2 $\upmu$m. We expect statistical
fluctuations of up to 2 $\sigma$ in 5\% of these points. However, given
their significance, such random fluctuations can have a large effect on the retrieval and
interpretation, and this should be borne in mind.

Finally, the NICMOS points for HD 189733b are not well reproduced by
any model within our suite. The opacity in this spectral region is
largely provided by collision-induced absorption of H$_2$ and He, so
it is difficult to think of a scenario in which this opacity could be
removed. The HD 189733b models already have significant cloud opacity
with a high top pressure, so a scenario in
which the opacity at shorter wavelengths could be increased
sufficiently to allow a fit to all of the STIS, ACS, WFC3 and NICMOS
points seems unlikely. The discrepancy between the models we show here
and those presented by \citet{sing16} has been traced to a lack of
sufficient collision-induced absorption in the models shown by
\citet{sing16} (Fortney 2015, private communication). The shortest
wavelength STIS points are also not well represented in some cases,
with the models appearing to lack sufficient absorption at these
wavelengths. We note that while we have assumed a uniform temperature for the retrieval, detailed analysis of the sodium line has revealed a hot thermosphere at the upper atmospheric layers for this planet (\citep{huitson12,wyttenbach15}.

\begin{figure*}
\centering
\includegraphics[width=0.85\textwidth]{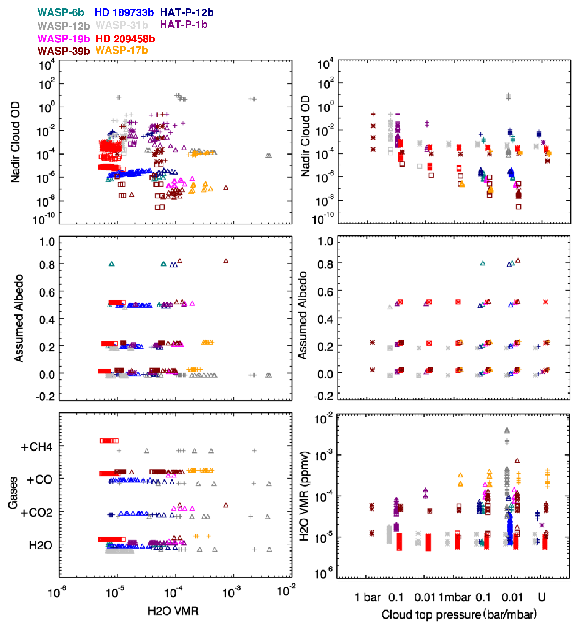}
\caption{Model parameters, where normalised ln($\chi^2_r$)$<$0.05. Colours correspond to the labels for each planet in
  Figure~\ref{specfits}. Symbols correspond to the different flavours
  of cloud model; crosses are extended Rayleigh scattering models,
  asterisks are extended grey models, diamonds are clear atmosphere
  models, open triangles are decade-confined Rayleigh models
  and open squares are decade-confined grey models. The
  different plots highlight correlations between different model
  properties. Points are shifted slightly for each planet for
  quantities with discrete values, such as cloud top pressure. \label{resultsplot}}
\end{figure*}

\begin{figure*}
\centering
\includegraphics[width=0.85\textwidth]{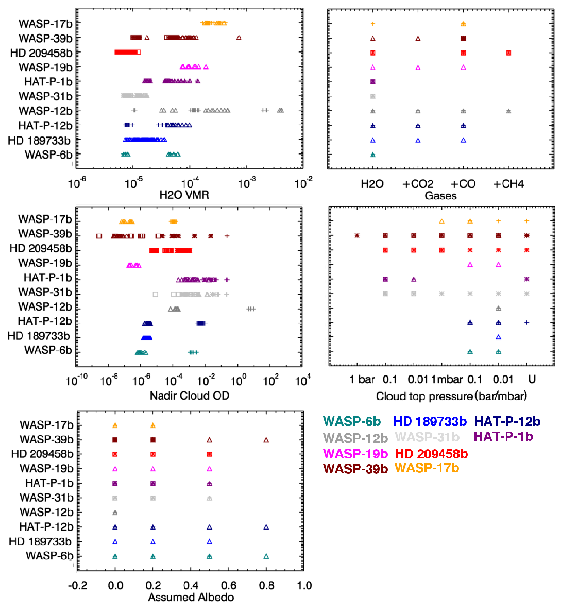}
\caption{Model parameters, where normalised ln($\chi^2_r$)$<$0.05, by planet. Symbols correspond to the different flavours
  of cloud model; crosses are extended Rayleigh scattering models,
  asterisks are extended grey models, diamonds are clear atmosphere
  models, open triangles are decade-confined Rayleigh models
  and open squares are decade-confined grey models. The
  different plots highlight correlations between different model
  properties. Points are shifted slightly for each planet for
  quantities with discrete values, such as cloud top pressure. \label{resultsplot_by_planet}}
\end{figure*}

The properties of the best fitting models are shown in
Figures~\ref{resultsplot} and~\ref{resultsplot_by_planet}. Each panel
in Figure~\ref{resultsplot} explores the parameter space
between two model variables to highlight any correlation between
properties. Variables with a lot of scatter are indications
that there is limited constraint from the
spectrum. Figure~\ref{resultsplot_by_planet} shows best fitting model
properties by planet. 

\subsection{H$_2$O and Other Molecular Absorbers}
H$_2$O abundances cluster between 0.01$\times$solar and solar values. HD
209458b has the lowest abundance, with between 5 and 10 ppmv. WASP-17b
has the highest constrained abundance, between 100 and 600 ppmv. The
solar value is approximately 500 ppmv. No constraint on H$_2$O
abundance is obtained for WASP-12b, as no water feature is clearly visible in
the spectrum, and for some models the cloud optical depth is so high that none would be
visible. For WASP-6b and WASP-39b, for which there is no WFC3
measurement, only a rough upper limit on H$_2$O abundance is obtained,
with this limit emerging from the facts that
no H$_2$O absorption features are observed in the long wavelength part
of the STIS spectrum, and the Spitzer levels imply molecular absorption.

The presence or absence of other molecular species can be constrained to some
extent by the relative transit depths at the wavelengths probed by
\textit{Spitzer}/IRAC at 3.6 and 4.5 $\upmu$m. However, due to
  the degeneracy inherent in using two data points to provide
  information about three gases, it is impossible to place any limits
  on the abundances of these gases. Figure~\ref{gases}
demonstrates that CO$_2$ would have low absorption at 3.6 $\upmu$m
but strong absorption at 4.5 $\upmu$m, whereas the opposite would be
true for CH$_4$. For H$_2$O only the 4.5 $\upmu$m absorption is
slightly stronger than 3.6$\upmu$m, with this effect slightly further pronounced
for CO. Because the effect of CO is small its presence is hard to rule
out, but CO$_2$ and CH$_4$ are more straightforward.

WASP-31b, WASP-6b and HAT-P-1b are fit best by models with H$_2$O
only. WASP-12b does not have strong constraints on the
presence of other gases. WASP-17b is best matched by models with either
H$_2$O only or H$_2$O plus CO. HD 209458b can be fit by any models
except those containing CO$_2$. HD 189733b, WASP-19b, WASP-39b and HAT-P-12b can be fit well with any model
except those containing CH$_4$. In aggregate, these findings are compatible with the likelihood of hot
planetary atmospheres having CO-dominated carbon chemistry over
CH$_4$, as the majority of constraints are in favour of either the
presence of CO, a lack of any gas except H$_2$O (which, given the
small effect of CO, does not provide strong evidence to rule it out), or a lack of CH$_4$. 

\subsection{Clouds}
The clearest division between groups of planets lies in the Rayleigh
cloud and grey cloud planets. HD 209458b, WASP-31b and WASP-39b are
all fit best by grey models (asterisks or squares in Figure~\ref{resultsplot}), whereas the other planets are better
represented by Rayleigh scattering cloud (crosses or
triangles). HAT-P-1b is the only planet with roughly equal sets of solutions for both
grey and Rayleigh cases, although WASP-31b and WASP-39b both have a
handful of Rayleigh solutions as well as grey solutions. No
planet is fit well with a completely clear atmosphere model, except
for a small minority of models for WASP-39b (diamonds). 

Cloud top pressures are tightly constrained for some planets, but not
for all. HD 189733b , HAT-P-12b, WASP-6b and WASP-12b have clouds with a top
pressure of 0.1 mbar or below, indicating that there is strong
evidence for cloud high in the atmosphere. Unsurprisingly, these
planets all also have results strongly in favour of Rayleigh
scattering clouds, which makes intiutive sense as it is easier to loft
small particles to high altitudes within an atmosphere. WASP-17b and
WASP-19b, the other two planets with solutions favouring Rayleigh
cloud, have cloud top pressures of 1 mbar or lower. HAT-P-1b has the
deepest cloud, with top pressures between 0.1 and 0.01 bar, consistent
with either grey or Rayleigh cloud.

Planets which have mostly grey cloud solutions have much broader
ranges of acceptable cloud top pressures. HD 209458b has a cloud top
range between 0.01 bar and below 0.01 mbar. The lower cloud top
pressures make less intuitive sense in this case as a grey cloud must
contain some large particles and these would be unlikely to be lofted very high,
esepcially as the optical depth for HD 209458b is in the middle of the
range spanned by all planets. WASP-31b has a similar range and optical
depth. WASP-39b has a slightly more logical correlation between cloud
top pressure and optical depth, with a low optical depth family of
solutions for 0.1 and 0.01 mbar cloud top pressures and a higher
optical depth set of solutions for 1 and 0.1 bar cloud tops. 

HD 189733b has strong evidence in favour of a decade-confined Rayleigh
scattering cloud. If this is a vertically confined aerosol layer high
in the atmosphere, it would suggest either a species that condenses at
cool temperatures only found high up, or a photochemically produced
haze that is short-lived deeper in the atmosphere. It is not known
what kinds of photochemical
products that could form and be stable at HD 189733b
temperatures.

One trend we uncover is that the coolest (WASP-6b, HAT-P-12b, HD
189733b, $T_{\mathrm{eq}}<$1300 K) and hottest
(WASP-12b, WASP-17b, WASP-19b, $T_{\mathrm{eq}}>$1700K) planets are the ones that have good evidence for
relatively high Rayleigh scattering aerosols (Table~\ref{temp_cloud}). WASP-39b is also
relatively cool but grey cloud models are favoured. The intermediate
temperature planets (HD 209458b and WASP-31b, with 1300
K$<T_{\mathrm{eq}}<$1700 K) are all also
best fit by grey cloud models. HAT-P-1b is fit best by Rayleigh
scattering models, but only deep clouds are favoured, so there is
little sensitivity to the scattering properties of the cloud. This
suggests that a clearing of cloud may occur at this temperature. Taking the cloud top pressures that
occur most frequently within the best-fitting models, the
planets with grey cloud solutions (except WASP-39b) are more likely to have deeper
cloud, which makes intuitive sense as larger particles are less likely
to be supported higher up in the atmosphere.

\begin{figure}
\centering
\includegraphics[width=0.5\textwidth]{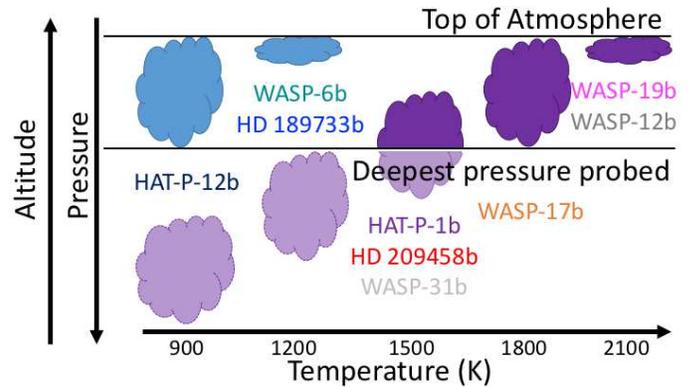}
\caption{A schematic illustrating how cloud structure for these hot
  Jupiters may vary with temperature, inferred from our retrieval
  results. For the hottest planets in the sample, WASP-19b and
  WASP-12b, the condensate we see can only exist relatively high in
  the atmosphere. For slightly cooler planets such as WASP-17b the cloud
  that originally forms high up in the atmosphere can extend
  downwards. Eventually, the cloud particles become large enough to
  sediment out and the cloud is only seen deep in the atmosphere
  (WASP-33b, HD 209458b, HAT-P-1b). For
  even cooler atmospheres, new species start to condense out and the
  sequence is repeated (WASP-6b, HD 189733b, HAT-P-12b)\label{cloudschema}}
\end{figure}

This may possibly indicate a continuum
of cloud formation through different mechanisms, or from different
substances. We show a schematic outlining these trends in
Figure~\ref{cloudschema}. As atmospheres cool, a cloud formed from the
same condensate will gradually fall deeper in the atmosphere, and
eventually new species will condense out forming new clouds in the
upper regions of the atmosphere. A similar sequence has been
postulated for brown dwarfs \citep{lodders06}. This sequence can explain both the trends
in scattering properties and the trends in cloud structure that we see
as a function of temperature. In particular, note that of the hot and
cold groups of planets with Rayleigh scattering cloud high in the atmosphere, WASP-6b and HD
189733b, and WASP-19b and WASP-12b, all favour models where the cloud
is confined to a limited pressure range in the atmosphere, whereas
WASP-17b and HAT-P-12b both favour extended cloud models.

At cooler temperatures near $\sim$1000K, the expected condensible
species of MnS, Na$_2$S and KCl are all highly scattering, while at
temperatures near 1500 K iron clouds could form which may be more grey
\citep{wakeford15}. Further observation will be necessary to test the effect
of stellar proximity, and therefore temperature, on hot Jupiter cloud
formation. An important question is whether other hot Jupiters also
follow this trend, which will be the subject of future work.

\begin{table} \centering \begin{tabular}[c]{|c|c|c|c|} 
\hline
Planet & $T_{\mathrm{eq}}$ (K) & Rayleigh/Grey & $P_{\mathrm{top}}$ (mbar)\\
\hline
HAT-P-12b &	963 & R & 0.01\\
WASP-39b & 1117 & R/G & 0.01\\
WASP-6b	& 1145 & R & 0.01\\
HD 189733b & 1201 & R & 0.01\\
HAT-P-1b & 1322 & R & 100\\
HD 209458b & 1448 & G & 10\\
WASP-31b & 1575 & G & 100\\
WASP-17b & 1738 & R & Top\\
WASP-19b & 2050 & R & 0.01\\
WASP-12b & 2530 & R & 0.01\\
\hline
\end{tabular}
\caption{We present evidence of a possible relationship between
                           equilibrium temperature and cloud
                           properties for this family of hot
                           Jupiters. Equilibrium temperatures are
                           taken from \citet{kataria16} except for
                           that of
                           WASP-12b, which is calculated for this work
                           using system values provided
                           by \citet{sing16}. The best-fit $P_{\mathrm{top}}$
                           quoted is the most commonly occurring value
                           within the
                           models where normalised ln($\chi^2<0.05$).
\label{temp_cloud}} 
\end{table}

\citet{stevenson16} find correlation between muted water
vapour features in WFC3 spectra and the planet's location in
temperature-log(g) space. Planets with higher equilibrium temperature
or log(g) are found to have stronger water vapour features than those
with low temperature and log(g), which is interpreted as a greater
likelihood of obscuring cloud in cooler/puffier planets. This makes
intuitive sense, as clouds are more likely to condense in cooler
atmospheres and less likely to sediment out in lower gravity
atmospheres.

\begin{figure}
\centering
\includegraphics[width=0.5\textwidth]{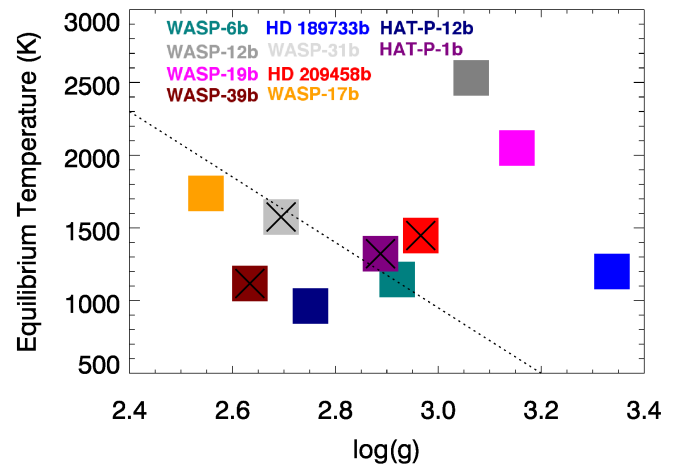}
\caption{The 10 planets in this study shown in effective
  temperature-log(g) parameter space. The dashed line indicates the
  demarcation between weak and strong H$_2$O features, which is
  suggested as a possible proxy for cloudiness by
  \cite{stevenson16}. The squares with crosses marked are those from
  our sample for which clear/deep cloud models provide the best
  fit. Uncrossed squares are planets for which high cloud is
  favoured, although for WASP-17b and WASP-19b this cloud is likely
  to be optically thin. \label{teff_logg}}
\end{figure}

We reproduce Figure 2 from \citet{stevenson16} with data for the 10
planets in this study (Figure~\ref{teff_logg}). The dashed line indicates the demarcation
between weak/strong H$_2$O features as found by
\cite{stevenson16}. However, we do not find any correlation between
position in this parameter space and the presence or absence of clouds
high in the atmosphere. This may be a result of the small sample sizes
used in each case, and also the fact that \citet{stevenson16} include
small planets in the sample whereas we only study hot Jupiters. 

The lack of agreement in our results may in part be due to differences
in the reduction of data, which is particularly notable for HD
189733b. This difference is discussed in more detail in
Section~\ref{hd189stuff}. However, this also suggests that data in
the STIS wavelength range is important for discriminating between
clear and cloudy atmospheres, as well as the WFC3 region.

\subsection{Temperature}
The assumed albedo, a proxy for temperature (zero
corresponds to zero dayside albedo and the same temperature at the
terminator as on the dayside) is poorly constrained for most planets,
as the temperature has a relatively small effect on the shape of the
spectrum, which is degenerate with the radius at the 10-bar pressure
level. However, for all planets except HAT-P-12b and WASP-6b, an albedo proxy
of 0.8 or higher is ruled out. For WASP-12b, WASP-31b, WASP-17b and
WASP-39b proxies of 0.5 or higher are ruled out. This generally
favours low albedo and efficient recirculation.

\subsection{Comparison With Sing et al.}
\citet{sing16} use empirical spectral indices, including the size of
the H$_2$O feature and the near-IR to mid-IR slope, to form an
initial categorization of the ten planets. In general, the trends uncovered
are borne out by this retrieval analysis. However, whereas
\citet{sing16} do not find evidence for any trend with temperature,
our more detailed analysis is able to show that the structure, and
probably the composition, of the cloud on these planets changes with increasing temperature. 

WASP-19b and WASP-17b are labelled as the clearest atmosphere planets
by \citet{sing16}. In our analysis, both planets have Rayleigh
scattering aerosol that can exist to quite low pressures, but higher
H$_2$O abundances than the majority of planets. These planets do not
have completely clear atmospheres, but the aerosol that is present
does not impact our ability to identify infrared molecular
absorption. There is an important distinction to be drawn between hazy
planets like this, and those with more opaque cloud such as HD
189733b.

There is some correlation between the type of cloud that provides the
best fit and the cloud coverage ordering presented by
\citet{sing16}. The planets \citet{sing16} determined to be the cloudiest -- WASP-6b, HD 189733b,
HAT-P-12b and WASP-12b -- are all
best represented in our analysis by Rayleigh scattering clouds with
relatively low top pressures. In general, and excepting WASP-19b and WASP-17b, the planets categorised as
less cloudy by \citet{sing16} are more likely to be represented by
grey cloud models with flatter spectra. This makes sense, as one of
the key spectral indices in determining cloudiness is the overall
slope from the optical to the mid-infrared, which is stronger for
Rayleigh scattering cloud than for grey. \citet{sing16} also finds
that HD 209458b, and to some extent WASP-31b, lie closer to their `cloudy' models (grey cloud
parameterisation) instead of the `hazy' models (Rayleigh scattering
cloud parameterisation). WASP-39b, another grey-cloud planet in our
analysis, also has a very shallow downward or possibly even upward
slope from the optical to the mid-infrared. These indices appear to be
fairly 
robust discriminators of grey versus Rayleigh scattering aerosols and can determine which planets will have large IR molecular features in their transmission spectra, but
degeneracies make it difficult to use similar techniques to rule out
the presence of aerosol altogether. 

We do not show results for Na and K abundances, as although these were
retrieved results were generally inconclusive. We also found some
difficulty in fitting the precise shape of these bands, as the centres
of the features are very narrow and the absorption tables used have a
limiting resolving power of 100 at 500 nm. Retrieved abundances of
CO$_2$, CO and CH$_4$ are also not presented, as precise constraints
were not obtained. 

\section{DISCUSSION}
Our ability to fit the spectra of ten very different hot Jupiters
using the same basic model demonstrates the power of spectral
retrievals to explore the atmospheres of transiting
exoplanets. However, it is clear that there is still significant
degeneracy within the dataset, and a good fit is harder to achieve for
some planets than others. Here, we briefly discuss the cases for which
a more planet-specific approach might be supposed to yield a better
fit quality.

\subsection{WASP-31b}
WASP-31b is the only candidate planet in the sample that is fit best
by \citet{sing16} with a multi-modal cloud model. A two component
cloud provides a Rayleigh scattering slope at short wavelengths and a
flat spectrum at longer wavelengths. The model set used in our
investigation did not include multi-modal clouds, although the
presence of such a cloud is plausible - Venus provides the best solar
system example (e.g. \citealt{knol82}). 

We test the effect of introducing a multi-modal cloud for retrievals
of WASP-31b. We include both a grey `cloud' and a Rayleigh scattering
`haze'. Haze top pressures of 0.01 mbar, 0.001 mbar and a uniform haze
were tested, and for each of these, cloud top pressures up to 0.01
mbar were tested (except for the 0.01 mbar haze top case, for which
the cloud only extends up to 0.1 mbar). 

\begin{figure}
\centering
\includegraphics[width=0.5\textwidth]{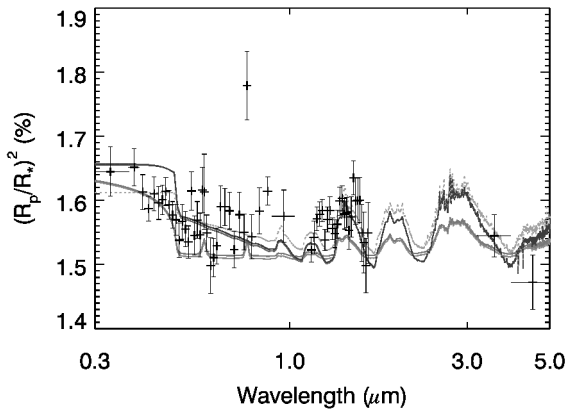}
\caption{WASP-31b spectrum with 2-component cloud models. Solid lines
  show haze with a top pressure of 0.01 mbar, dotted lines show has
  with a top pressure of 0.001 mbar. Darker shades correspond to
  lower cloud top pressures. None of these models produce a
  substantially better fit to the spectrum.\label{wasp31_plots}}
\end{figure}

Although the apparent shape of the spectrum seems to favour this kind
of two-component model, the reduced $\chi^2$ values do not provide any
strong evidence for this over a simpler model. Two-component cloud
model plots are shown in Figure~\ref{wasp31_plots}, with the lowest
reduced-$\chi^2$ values actually slightly higher than those of the best-fitting
single-cloud models.

\subsection{HD 189733b}
\label{hd189stuff}
The data for HD 189733b are probably the most constraining of the ten
planets. However, there are some clear discrepancies between the
models and the observed spectrum. The most obvious is the very low
transit depth for the two NICMOS points in the 2 $\upmu$m region. 

\begin{figure}
\centering
\includegraphics[width=0.5\textwidth]{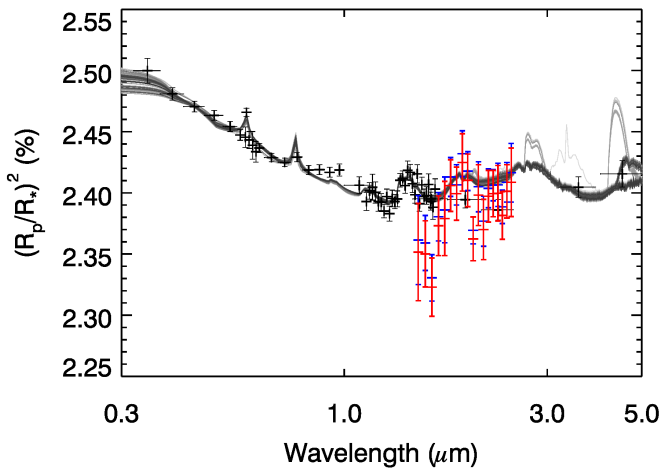}
\caption{HD 189733b model spectra, plus data including the full
  NICMOS spectrum from \citet{gibson12} (blue) and as modified by
  \citet{pont13} (red). There is an obvious mismatch with the WFC3
  spectrum and models at the bluer end of the NICMOS range, but moderate
  agreement with models elsewhere due to the large error bars
  on the spectrum. The broad spectral shape is consistent with
  absorption by H$_2$O + H$_2$-H$_2$ collision-induced absorption, as
  shown in the models.\label{nicmos}}
\end{figure}

The binned NICMOS points used in this paper are as presented
by \citet{pont13}, but they were binned up from a spectrum that was
originally reduced by \citet{gibson12}. This original
spectrum has large error bars and also a large amount of scatter. If
the original spectrum from \citet{gibson12} is overplotted (Figure~\ref{nicmos}) it can be
seen that the majority of points longwards of 1.6 $\upmu$m match the
models quite well. The averaged points have their transit depths
brought down by three low values at the shortest wavelength end, and
two outlying low points at longer wavelengths. The transit depths were
fractionally reduced by the starspot correction applied by
\citet{pont13}, and subsequently the models do not provide quite such a good
match to the data.  

It is impossible to fit the binned points with any plausible model
spectrum. In order to simultaneously fit the Rayleigh-dominated slope in the STIS and ACS
data and the small water feature, a fairly low H$_2$O abundance is
required. This means that H$_2$-H$_2$ collision-induced absorption
features become visible between the water bands, providing a hard
opacity floor. Even if the spectrum could be fit with a combination of
Rayleigh cloud and grey cloud, as suggested for WASP-31b, which would
potentially allow H$_2$O abundances closer to the solar value, the opacity floor would still fall above the NICMOS points as the longer
wavelength H$_2$O features would then be stronger. 

Given the complicated systematics of the NICMOS instrument, discussed
at length by \citet{gibson12}, and especially in the light of the
obvious discrepancy with WFC3 results at the shorter wavelength end,
we conclude that the lack of a good match to these points does not
provide strong evidence of innacuraices in our best fit model. 

\subsubsection{Star Spots and Spectral Stitching}
\label{starspots}
Each of the spectra in this survey were compiled from segments of
spectrum obtained by different instruments at different times. As
discussed by \citet{pont13} in the context of HD 189733b, and
\citet{barstow15} with a view towards \textit{JWST}, changing starspot
coverage between the times when different datasets were obtained can
affect the accuracy of spectral stitching. \citet{pont13} describe
star spot corrections for stitching of HD 189733b, based on
ground-based monitoring of the stellar flux over several
years. Individual spectral points are shifted up or down depending on the
relative numbers of occulted and unocculted spots estimated to be
present at the time of observation, as especially in the visible the
effect of starspots increases towards shorter wavelengths. A strong downward
slope from visible wavelengths through to the infrared remains even
after correction for starspots.

\citet{mccullough14} postulate that it is possible to explain the broad
spectral shape of the HD 189733b observations with starspots alone,
without invoking the presence of clouds, and under this assumption
they are able to fit the spectrum with a clear solar composition
atmosphere model. Their hypothesis assumes a more
significant effect from unocculted star spots (spots outside the
transit chord) compared with occulted spots, and also a more
  substantial effect that that estimated by \citet{pont13}, which was
  arrived at by monitoring the star and analyzing occulted starspots
  in transit lightcurves. This is a difficult question to resolve
since the spot coverage when HD 189733 is at its brightest is an
unknown quantity. 

However, HD 189733b is
the most active star in the dataset considered here. A common proxy
for stellar activity is the log[$Ca_{HK}$] value, which is a measure of
the emission in the H and K Fraunhofer lines from singly-ionised
calcium. 
HD 189733b has a log[$Ca_{HK}$] value of -4.501, higher than
the values for the other stars in this study; despite this, other
transmission spectra
display slopes of similar magnitude in the visible, including those of
planets such
WASP-12b and HAT-P-12b which have log[$Ca_{HK}$] values of less than
-5. This, combined with the fact that very substantial spot
  coverage needs to be invoked to explain the slope for HD 189733b,  would
indicate that the presence of aerosol is a more likely explanation for
these slopes. We believe that the evidence from HD 189733b
  indicates that spots could probably not reproduce this same effect across planets orbiting
less active (and by extension less spotty) stars.

\subsection{WASP-12b}
As discussed by \citet{sing13}, the WASP-12b transmission spectrum is
extremely challenging to fit with Rayleigh scattering cloud models. The best fit obtained
by \citet{sing13} over a range of atmospheric temperatures is for Mie
scattering Al$_2$O$_3$. Fits using Rayleigh scattering particles are
hampered by the requirement for low atmospheric temperatures of
around 800 K, which seems to be incompatible with WASP-12b's expected
high equilibrium temperature.  

We find that the
best fits from our limited model set are provided by an optically thick Rayleigh
scattering aerosol with a 0.01 mbar top pressure, but the quality
of fit decreases towards shorter wavelengths, with more opacity
required than is provided by any of the models. Rather than finding a
trend towards cooler atmospheric temperatures, we actually find a fit
consistent with a zero dayside albedo and strong recirculation - hot
atmospheres are strongly favoured for this planet. It is likely that
more complex Mie scattering clouds may provide a better fit to the
observed spectrum, but there would be significant
degeneracy between various cloud parameters, so a detailed exploration
will most likely be deferred until after observation with
\textit{JWST}. 

\citet{kreidberg15} recently published a new WFC3 transmission
spectrum for WASP-12b, with increased precision. These results
show a clear H$_2$O detection. Analysis in the paper, including with
the NEMESIS code, found H$_2$O volume mixing ratios between 10 and
10,000 ppmv, and analysis with the CHIMERA code \citep{line13} ruled out an
atmosphere with C:O $>$1 at 3$\sigma$ confidence. 

\begin{figure}
\centering
\includegraphics[width=0.5\textwidth]{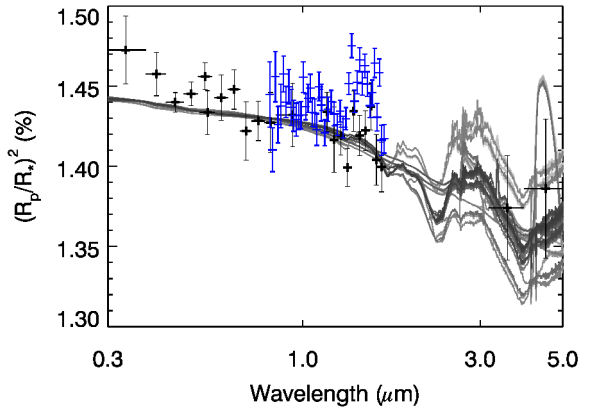}
\caption{WASP-12b spectra including the newer WFC3 spectrum from
  \citet{kreidberg15}. There is an obvious offset between this
  spectrum and the existing data, resulting from differences in the system parameters and limb darkening used in the analysis.
This is a
  common
  issue in transit spectroscopy and was discussed by
  \citet{kreidberg15}.\label{newwfc3}}
\end{figure}

We do not attempt to perform a retrieval with these data due to the
difficulty of combining different analysis using different limb darkening and system parameters. Figure~\ref{newwfc3} shows that there is a
clear offset between the new WFC3 data and the previously obtained
spectrum. In addition, all data we consider from \citet{sing16}
have been reduced using the same lightcurve analysis pipeline, and
this would be required for a rigorous analysis of the new WFC3 data in
the context of STIS and Spitzer measurements.However, it is
  clear from Figure~\ref{newwfc3} that the observed water vapour
  feature from Kreidberg et al. is larger than the feature predicted
  by any of our best-fit models. To fit this feature, models would
  need one or more of the following modifications: a substantially increased H$_2$O
  abundance; a reduced cloud optical depth; or a hotter upper
  atmosphere resulting in an increased scale height. As stated above,
  optically thick cloud appears to be required by the steep slope in
  the STIS measurement, and it would not be expected for such a hot
  planet to have a substantially super-solar H$_2$O abundance. The
  temperature of the upper atmosphere, to which thermal emission measurements are relatively
 insensitive \citep{stevenson14b}, is the least well-constrained
 property here. Future analyses with consistently reduced data and a
 greater range of temperature profiles may shed further light on this
 planet.

\subsection{Influence of Spitzer data}

The Spitzer data provide the only information at wavelengths $>$ 2
$\upmu$m in these retrievals, and as such they provide a critical role
in our conclusions about the presence of molecular species other than
H$_2$O. However, there has been considerable debate about the
reliability of Spitzer photometry in the past, with reanalyses
producing substantial shifts in transit depths (e.g. \citealt{diamond14,evans15}) and conclusions drawn
from only two points are obviously highly degenerate.

\begin{figure*}
\centering
\includegraphics[width=0.85\textwidth]{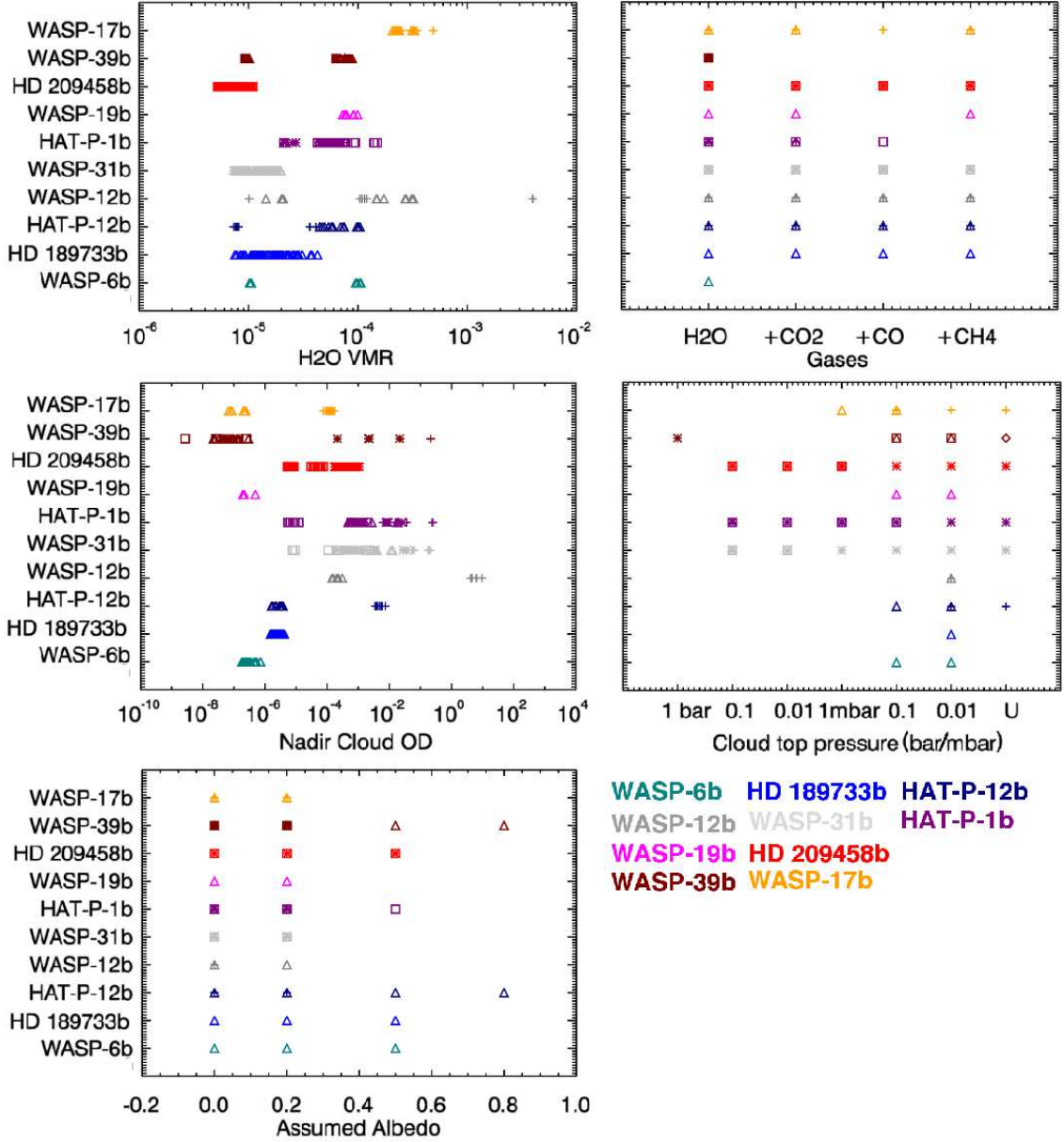}
\caption{Model parameters, where normalised ln($\chi^2_r$)$<$0.05, by
  planet, for retrievals without the \textit{Spitzer} data. Symbols correspond to the different flavours
  of cloud model; crosses are extended Rayleigh scattering models,
  asterisks are extended grey models, diamonds are clear atmosphere
  models, open triangles are decade-confined Rayleigh models
  and open squares are decade-confined grey models. The
  different plots highlight correlations between different model
  properties. Points are shifted slightly for each planet for
  quantities with discrete values, such as cloud top pressure. \label{nospitzer}}
\end{figure*}

We test the influence the \textit{Spitzer} data on our results by
running the retrievals again without the \textit{Spitzer} points. The results
are presented in Figure~\ref{nospitzer}. There are few substantial
differences between these results and the originals (Figure~\ref{resultsplot_by_planet}) including the
\textit{Spitzer} data. The key differences are 1) removing the \textit{Spitzer} data
for HAT-P-1b means that this planet is better fit by grey cloud models
than Rayleigh scattering models; and 2) a reduction in our
ability to draw conclusions about the presence of molecular species other
than H$_2$O. The second of these consequences is entirely to be
expected, since the only information we have about these other gases
comes from the Spitzer points. 

WASP-39b and WASP-6b favour H$_2$O-only models when the
  \textit{Spitzer} points are removed; however, these two planets now
  only have STIS data, meaning that no molecular species are
  detected at all. This is therefore simply a consequence of a penalty
  on the reduced-$\chi^2$ when an extra model parameter is
  included. In conclusion, the \textit{Spitzer} points do provide some
constraint for some planets on the presence of molecular species other
than H$_2$O, but have relatively little influence on any other
retrieved parameters, and therefore have little bearing on our
conclusions about the cloud properties of these worlds, with the
exception of grey versus Rayleigh scattering on HAT-P-1b.

\subsection{Comparison With Climate Models}

\citet{kataria16} present a General Circulation Model (GCM) survey of
nine of the ten hot Jupiters discussed here. WASP-12b is deferred to a
later study due to its high equilibrium temperature. This study is based on
the Substellar and Planetary Radiation and Circulation (SPARC) model
\citep{showman09}. The model is cloud-free and calculates temperature
structure, wind fields and gas mixing ratios across the planet. Here,
we compare the model predictions from \citet{kataria16} with the
retrieval results in this work. 

\subsubsection{Gas Abundances}
The gas abundances in the GCMs are derived from solar composition
(after \citealt{lodders03})
thermochemical equilibrium models. As all planets are hot, CO
chemistry is expected to generally dominate over CH$_4$ chemistry, so the two
most abudant molecular species after H$_2$ are H$_2$O and
CO. CO$_2$ has very low abundances for all planets and the CH$_4$
abundance decreases from potentially observable levels at the
teminators for the coolest
planets (HAT-P-12b, WASP-39b, WASP-6b and HD 189733b, with equilibrium
temperatures below 1300 K) to levels several orders of magnitude below
CO for the others. 

The spectral data available provide limited constraint on the presence
of gases except H$_2$O. However, the H$_2$O abundances from the
retrievals are sub-solar (by between 1 and 2 orders of
magnitude) for the majority of the ten planets in our
retrievals. WASP-17b is the exception, with approximately solar H$_2$O
(a volume mixing ratio of $\sim$5$\times$10$^{-4}$). 

Despite a generally lower H$_2$O volume mixing ratio than a solar
composition atmosphere, as mentioned above our results are consistent
with CO-dominated carbon chemistry. We see no evidence for CH$_4$
absorption for any planet, and the only planets for which any
solutions including CH$_4$ are possible are WASP-12b (which, although
not dealt with in the GCM study, is almost certainly too hot to have
substantial CH$_4$) and HD 209458b. With the caveat that
  constraints on the presence of CO$_2$, CO and CH$_4$ are not strong,
  and depend heavily
  on the \textit{Spitzer} points, this would suggest that CO
chemistry dominates on all planets in the sample, including the cooler
ones, and therefore disequilibrium effects may be important.

\subsubsection{Condensate Formation}
\citet{kataria16} also include estimates of which condensates are
likely to form at different pressure levels on each planet. This is of
particular interest for HD 189733b, which has strong evidence for a
vertically confined cloud. Based on these models, condensates that would only form at low
pressures could be KCl or ZnS (western terminator profile) or Na$_2$S
(eastern terminator profile). These particles have somewhat different
properties to MgSiO$_3$ (enstatite) which has previously been
considered as a possible condensate on HD 189733b. \citet{lee14} found
that Na$_2$S is consistent with the terminator spectrum of HD
189733b. 

Condensate cloud formation is very
heavily dependent on the spatially varying temperature profile and
also on circulation, for which we currently have very little
constraint. In addition, the majority of models simply consider the
condensation curve at equilibrium when calculating the height of cloud
condensation. Constraints on cloud structure such as those obtained
for HD 189733b can start to inform directions of research for
exoplanet cloud models, but the higher spectral quality of
\textit{JWST} will most likely be required for significant
advances. \citet{wakeford15} demonstrate how the increased spectral
resolution and coverage of \textit{JWST} spectra can further constrain
cloud properties.

\citet{wakeford16} summarise the condensation temperatures of a
variety of possible hot Jupiter cloud constituents, along with the
available masses of each condensate. For temperatures greater than
1700 K, the majority of possible condensates are Ti and Al
compounds, with Ti and Al being the limiting species for these cloud
constituents. Below this temperature, the condensates with the highest
available masses are Fe and and magnesium silicate minerals; these
available masses outnumber those for Al-based compounds by an order
of magnitude and Ti based compounds by a factor $\sim$100. The
theoretical condensation temperatures do not correspond exactly to the
temperatures where we observe transitions between the cloud types. However,
it is conceivable that the reason the hotter planets with extended
clouds (WASP-17b, WASP-19b and WASP-12b) tend to have lower cloud
optical depths and more visible molecular features than the colder planets
is simply that the available mass for the hotter condensate is lower
than that for the colder condensate.

\subsubsection{Hemispheric Asymmetry}
\citet{kataria16} predicts substantial hemispheric asymmetry in
temperature and chemistry for many of the planets in the sample. This
is a consequence of the strong eastward jets that are thought to occur
on tidally-locked hot Jupiters. Evidence for these jets is found in
the offset of thermal phase curve hot spots
(e.g. \citealt{knutson12,stevenson14b}) and is predicted by GCMs.

\citet{kataria16} find that in general the eastern terminator region
is relatively warm, but for all but the most highly irradiated
planets the western teminator is actually the coldest region on the
planet. Our retrieval models are 1-dimensional, so we are effectively
representing an average of two potentially very different limbs. These
large temperature differences between the limbs raise the possibility
of cloud material condensing out at different pressure levels on
different limbs. An extreme case would be one cloud-free limb and one
cloudy limb.

\citet{line16} explore the impact of this extreme scenario, and find
that partially cloudy models can produce very similar WFC3 spectra to
cloud-free models with a high mean molecular weight atmosphere. Given
the size and temperature of the objects we consider in this study it
is most likely that they are all H$_2$-He-dominated; however, it is
possible that their east and west limbs do have very different cloud
properties, as well as different temperatures. The combined effect of
different temperatures and different cloud coverage has yet to be
explored. 

The precisions of existing spectra are insufficient for distinguishing between patchy and global cloud
scenarios. As pointed out by \citet{line16}, the best method of
distinguishing is to examine a high signal-to-noise light curve for
asymmetries, which may become possible with JWST. 

\subsection{More Complex Cloud Models}
\citet{benneke15} perform a similar retrieval analysis of multiple hot
Jupiter spectra using the SCARLET model, which combines a retrieval
algorithm with physically motivated priors. Cloud models are also
somewhat more complex that those explored here, with cloud
parametrized according to particle size, condensate mole fraction, top
pressure and profile shape factor. This allows exploration of a
greater range of cloud models.

\citet{benneke15} discuss retrievals of eight hot Jupiter WFC3
spectra, including HD 189733b, HD 209458b and WASP-12b, for which STIS
measurements are also considered. \citet{benneke15} finds that HD
209458b and WASP-12b can both be fit well by a thick cloud with a top
below 0.01 mbar, and depending on the starspot correction used HD
189733b may have either a Rayleigh scattering cloud or a similar thick
cloud. 

The findings for the type of cloud on HD 209458b  are very consistent
with our results, although we retrieve somewhat sub-solar water vapour
abundances. These abundances are not ruled out by \citet{benneke15},
although low water abundances of 0.1$\times$ solar are only compatible
with cloud tops deeper than 100 mbar, whereas our deepest cloud top is
10 mbar. However, a direct comparison between the two models is
difficult as the parameterisations are very different; there is
significant degeneracy between the specified cloud top pressure and
the particle number density, as the two combine to determine the pressure at
which optical depth $\tau$=1.

\section{CONCLUSIONS}
We have performed retrievals for the ten hot Jupiter spectra presented
by \citet{sing16}, using a consistent set of models and cloud
parameterisations. The lack of planet-specific model adjustment,
except where justified by prior knowledge of bulk properties and
stellar irradiation, has enabled us to consistently examine the best fit
atmospheres across all ten planets. We find that all spectra are
consistent with at least some aerosol or cloud, and there is a clear
split betwen planets with clouds dominated by small,
Rayleigh-scattering particles and clouds dominated by larger particles
with grey spectral effects. 

The level of constraint obtained is highly variable across the ten
planets, due to differences in data quality and coverage. Strong
constraints on cloud properties are obtained for HD 189733b and HD
209458b, with HD 189733b requiring a vertically confined Rayleigh
scattering cloud layer at high altitudes and HD 209458b requiring a
lower altitude grey cloud. These two well-studied planets remain good
examples of two very different hot Jupiters. Unexpectedly, the very
hot planets WASP-17b, WASP-19b and WASP-12b all show strong evidence
for the presence of Rayleigh scattering clouds, mostly at relatively high altitudes. This
may be due to condensation becoming possible higher in the
atmosphere. We find that planets with $T_{\mathrm{eq}}<$1300 K or
$>$1700 K are likely to have Rayleigh scattering clouds made of small
particles, whereas planets with intermediate temperatures are likely
to have grey, deeper cloud.

Our use of an optimal estimation retrieval algorithm, which
  does not allow full marginalization over the available parameter
  space, has necessarily introduced some level of model dependence in
  these results, and our conclusions should be viewed with that in
  mind. However, as all planets have been treated in a precisely
  similar way, the comparisons between the broad cloud characteristics
  of each planet may be considered robust.

Large observation programmes, obtaining similar data products for a
range of planets, are rapidly proving to be powerful tools, enabling
comparative planetology for exotic exoplanet atmospheres. Future
observations with \textit{JWST} will no doubt shed further light on
these fascinating worlds. New hot Jupiters are still being discovered
and are gradually filling up the family tree, and the retrieval techniques
 that we have applied to this subset have revealed trends in cloud parameters that bear further
investigation in the future. 

\section*{ACKNOWLEDGEMENTS}
We thank the anonymous referee for their comments and suggestions. JKB
is currently funded under European Research Council project 617119 (ExoLights). JKB also
acknowledges the support of the Science and Technology Facilities
Council during earlier stages of this work. SA acknowledges the support of the
Leverhulme Trust (RPG-2012-661) and SA and PGJI receive funding from the Science and Technology
Facilities Council (ST/K00106X/1). DKS acknowledges support from the European Research Council under the European Union’s Seventh Framework Programme (FP7/2007-2013)/ERC grant agreement number 336792.  
 
This work is based on observations with the NASA/ESA HST, obtained at the Space Telescope Science Institute (STScI) operated by AURA, Inc. This work is also based in part on observations made with the Spitzer Space Telescope, which is operated by the Jet Propulsion Laboratory, California Institute of Technology under a contract with NASA. 
\bibliographystyle{apj}
\bibliography{bibliography}

\begin{figure*}
\centering
\includegraphics[width=0.85\textwidth]{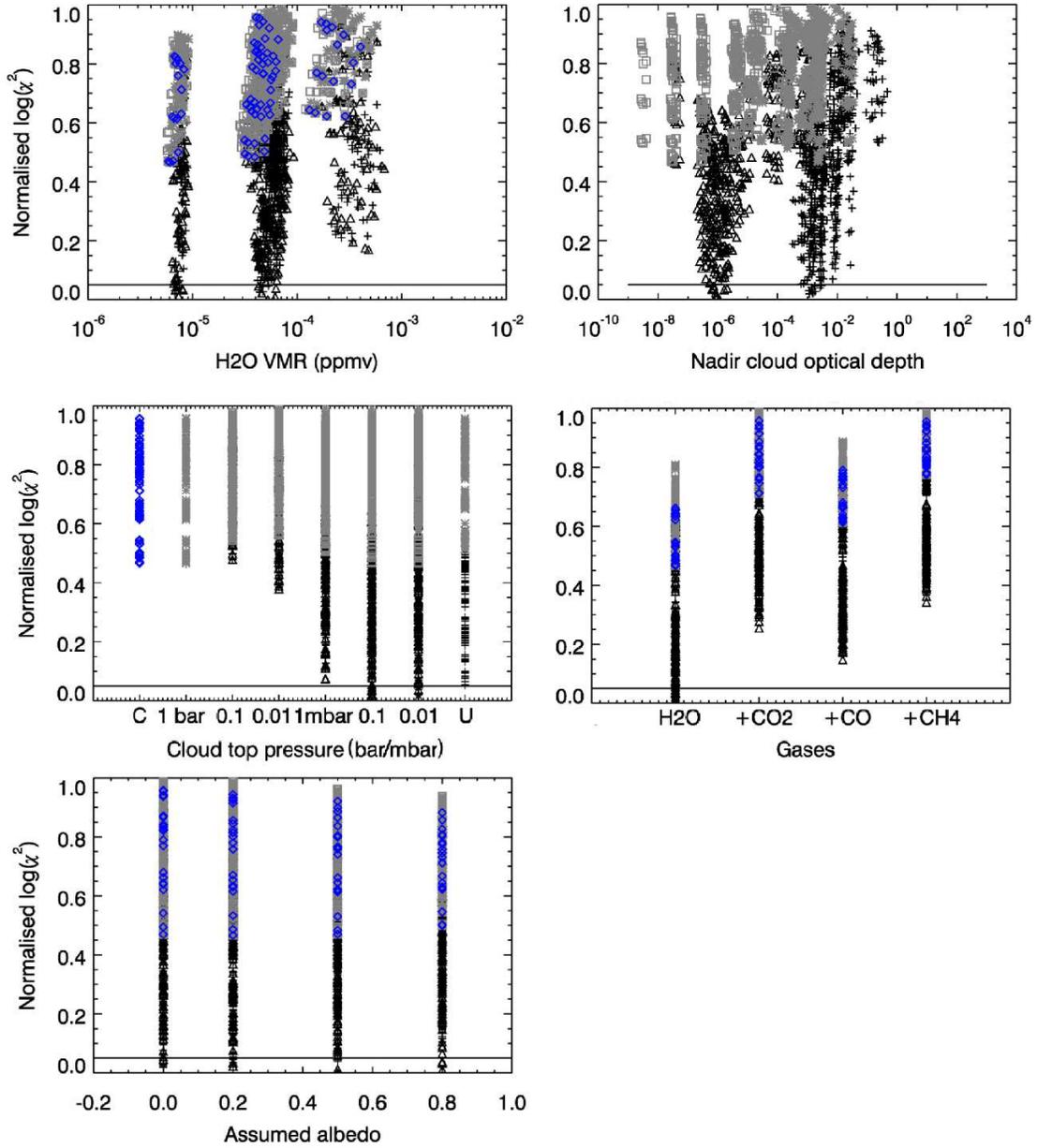}
\caption{Full results for WASP-6b for H$_2$O abundance, albedo proxy,
  cloud optical depth and top pressure, and molecular species
  present. Black crosses (triangles) are Rayleigh scattering extended
  (confined) cloud models, grey asterisks (squares) are grey extended
  (confined) models, and blue diamonds are clear atmosphere
  models. Each point corresponds to one run of 3,600. The horizontal line shows the cut off reduced-$\chi^2$ for
  the top 2 percent - all points below this were included in the
  results in the main body of the paper. WASP-6b has
  little constraint on H$_2$O abundance due to the lack of WFC3
  observation for this planet.\label{wasp6b}}
\end{figure*}

\begin{figure*}
\centering
\includegraphics[width=0.85\textwidth]{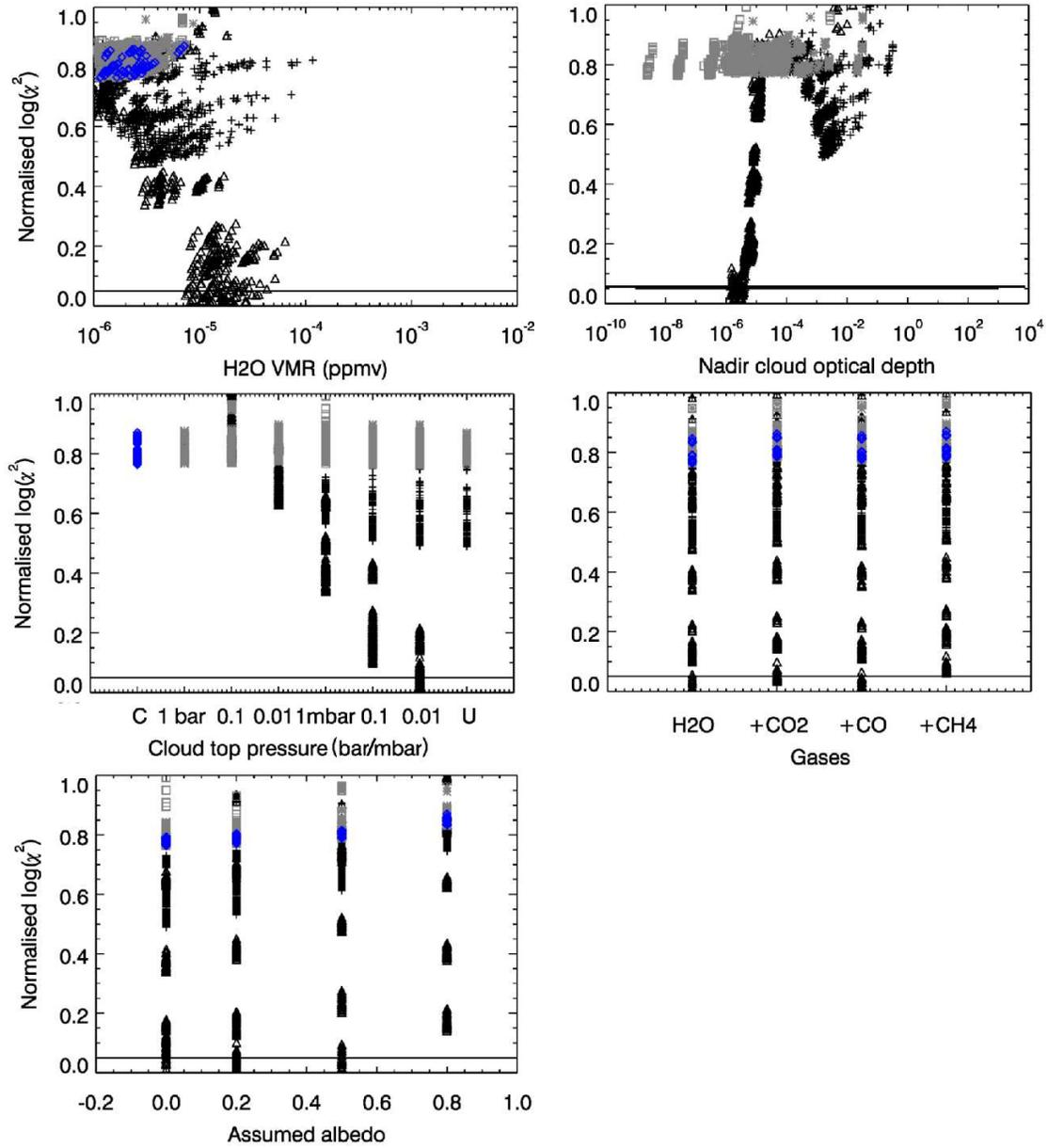}
\caption{Full results for HD 189733b, as Figure~\ref{wasp6b}.HD
  189733b has strong constraints on cloud top pressure and optical depth.\label{hd189733b}}
\end{figure*}

\begin{figure*}
\centering
\includegraphics[width=0.85\textwidth]{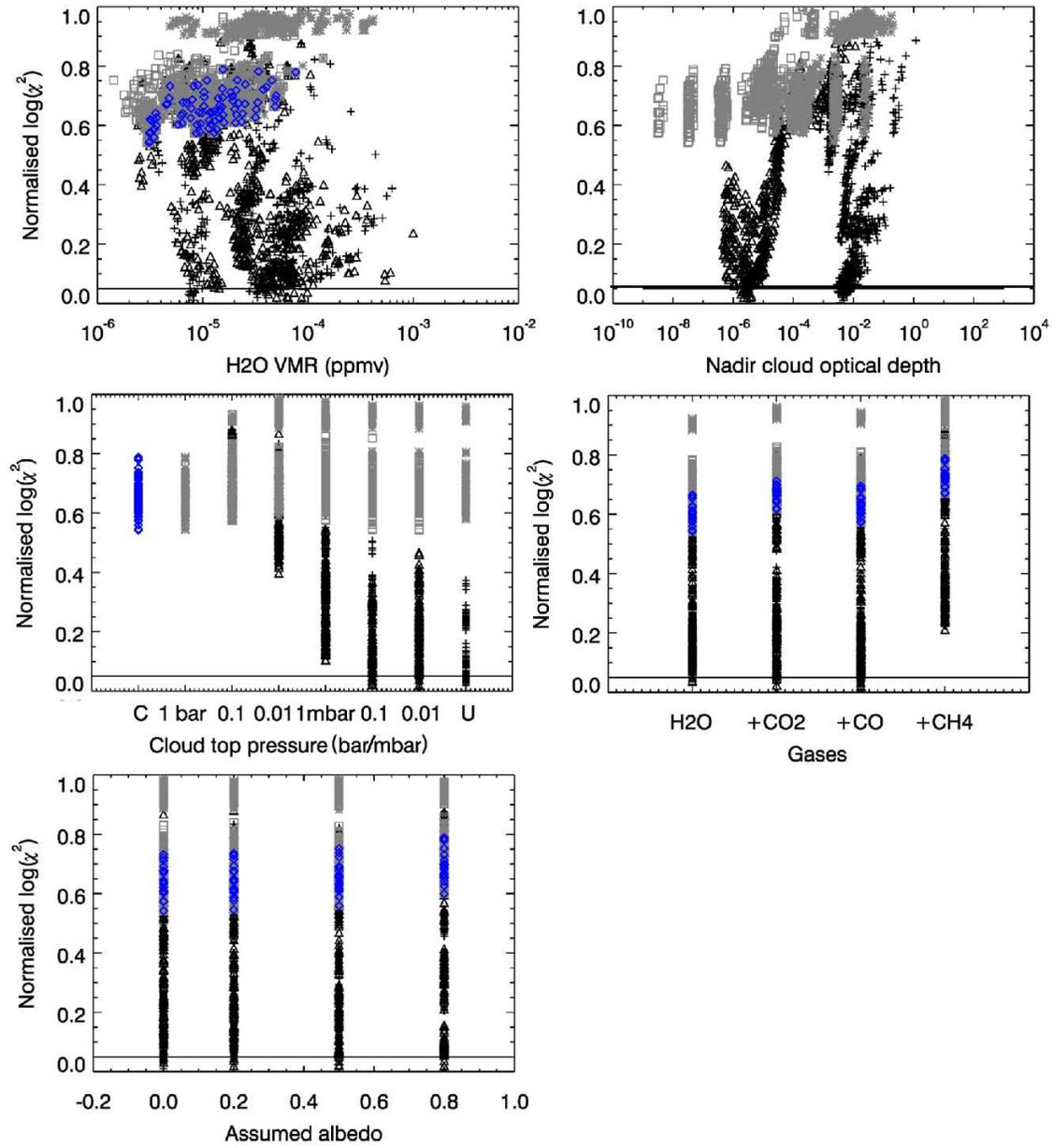}
\caption{Full results for HAT-P-12b, as
  Figure~\ref{wasp6b}. HAT-P-12b has two families of solutions with
  different cloud optical depths and H$_2$O abundances.\label{hatp12b}}
\end{figure*}

\begin{figure*}
\centering
\includegraphics[width=0.85\textwidth]{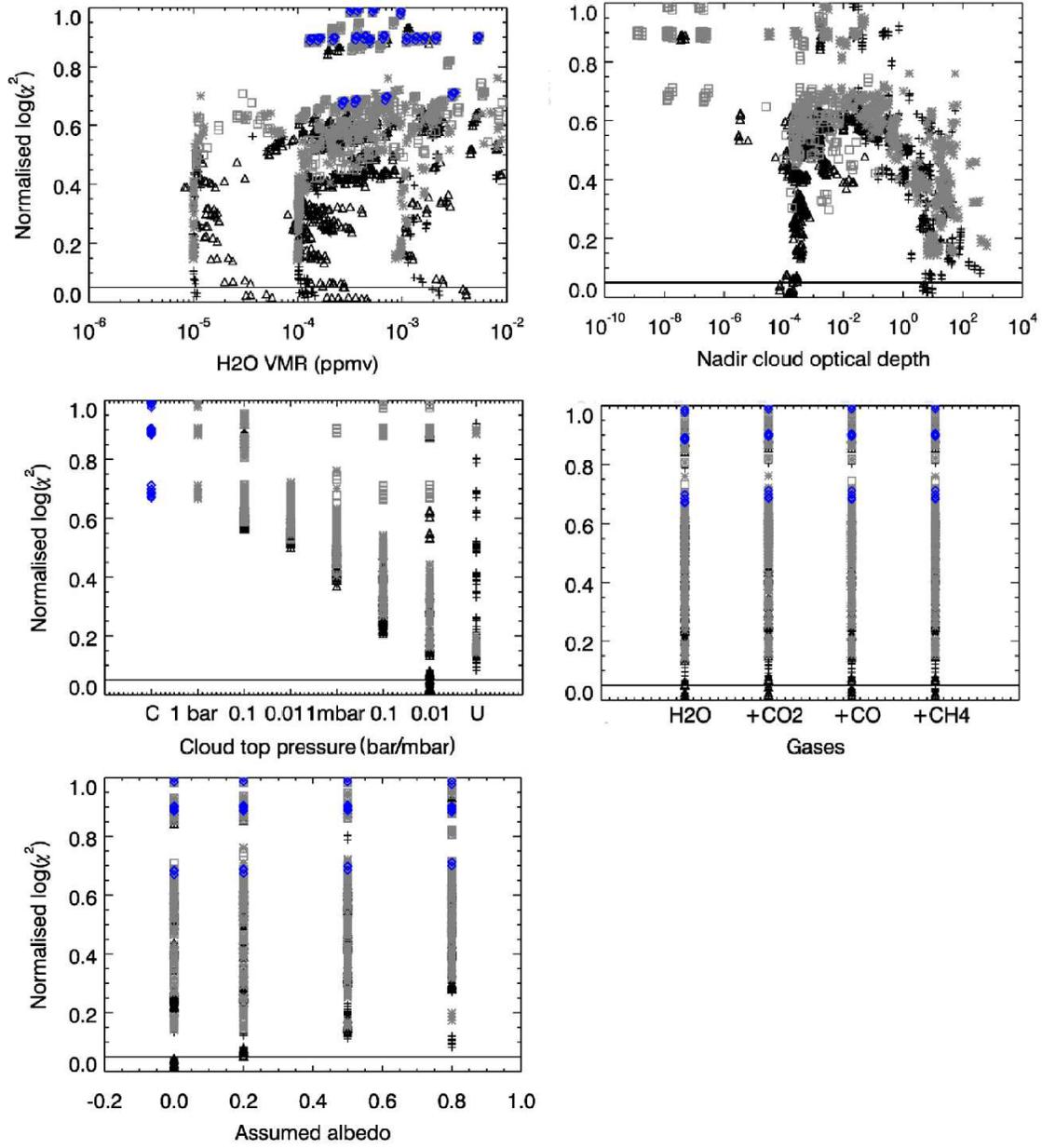}
\caption{Full results for WASP-12b, as Figure~\ref{wasp6b}. Due to
  the scatter on the WFC3 points no constraint on H$_2$O abundance is obtained.\label{wasp12b}}
\end{figure*}

\begin{figure*}
\centering
\includegraphics[width=0.85\textwidth]{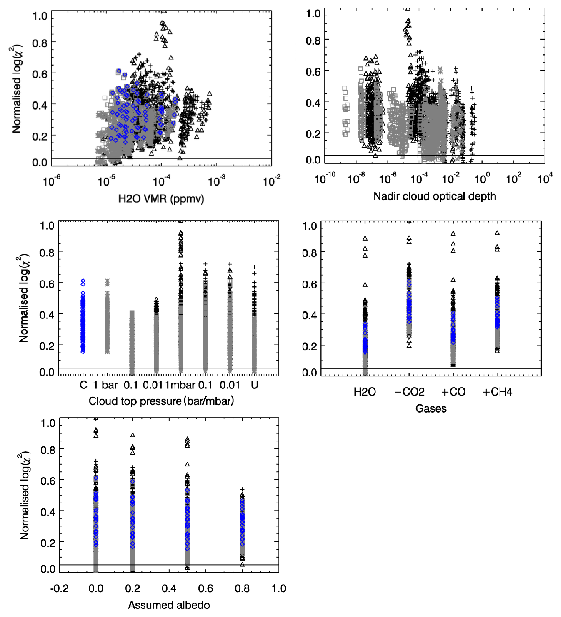}
\caption{Full results for WASP-31b, as Figure~\ref{wasp6b}.\label{wasp31b}}
\end{figure*}

\begin{figure*}
\centering
\includegraphics[width=0.85\textwidth]{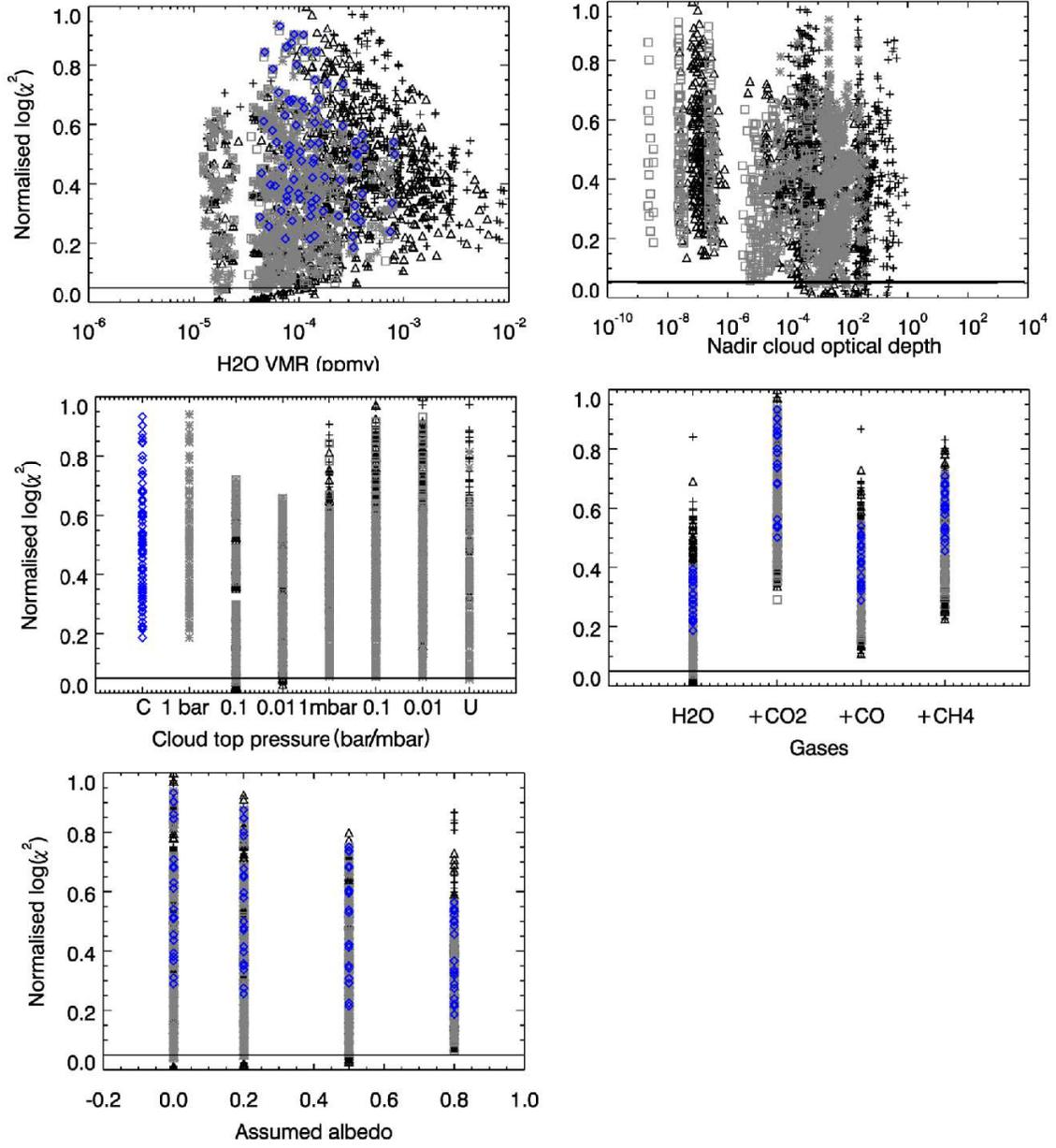}
\caption{Full results for HAT-P-1b, as Figure~\ref{wasp6b}. HAT-P-1b
  is the planet with the most equal distribution of grey and Rayleigh
  scattering models.\label{hatp1b}}
\end{figure*}

\begin{figure*}
\centering
\includegraphics[width=0.85\textwidth]{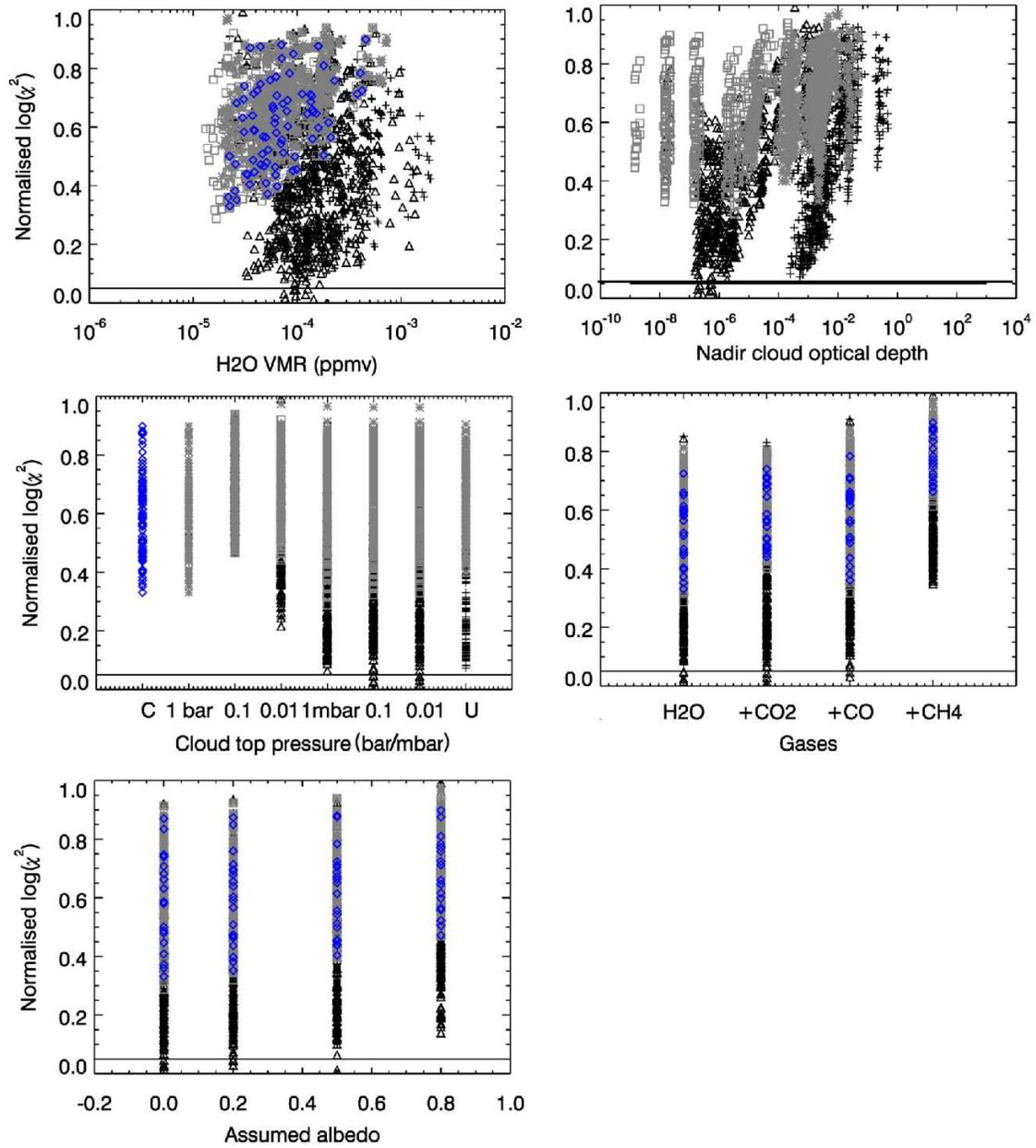}
\caption{Full results for WASP-19b, as Figure~\ref{wasp6b}. WASP-19b
  has a tight constraint on H$_2$O abundance.\label{wasp19b}}
\end{figure*}

\begin{figure*}
\centering
\includegraphics[width=0.85\textwidth]{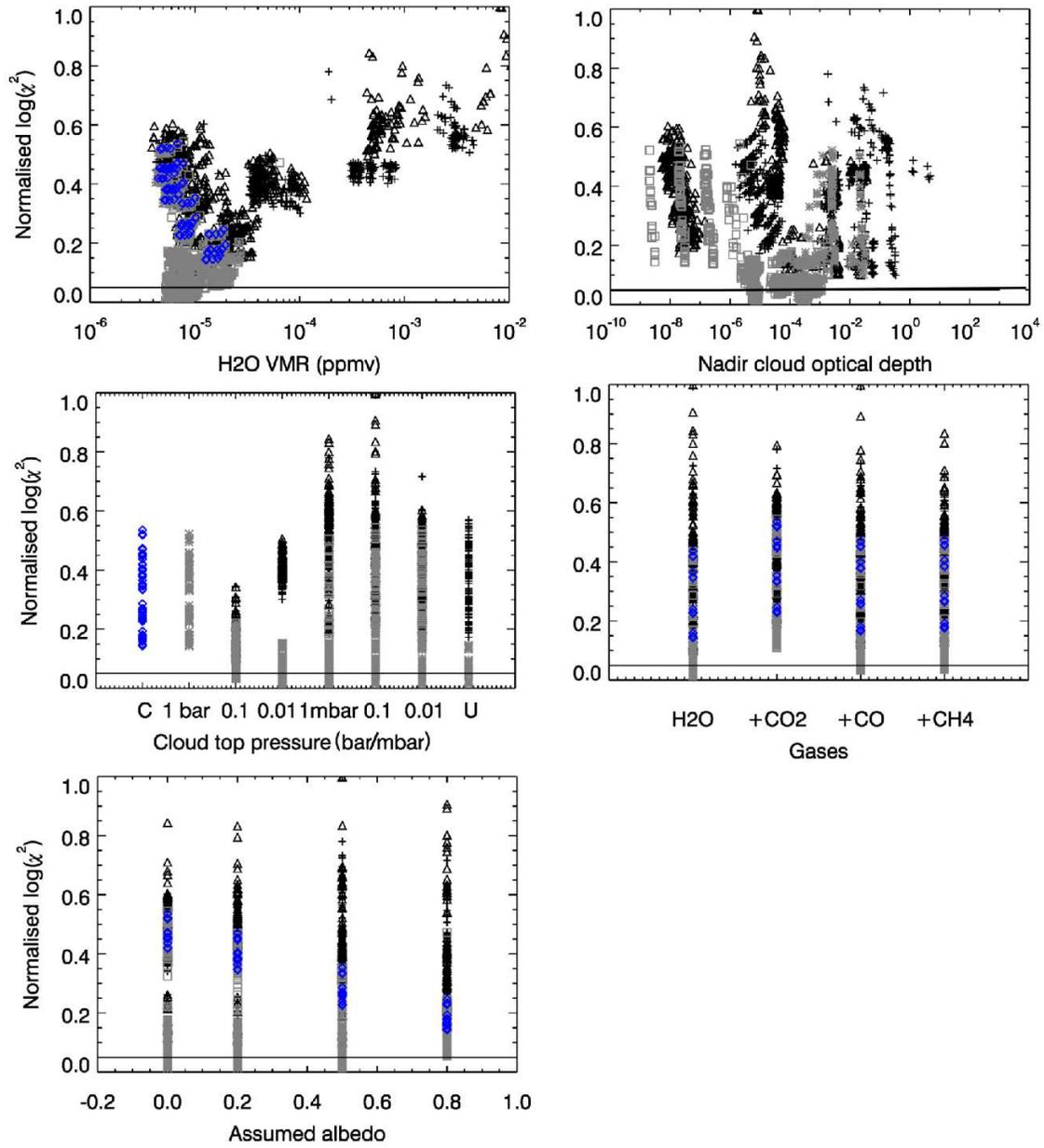}
\caption{Full results for HD 209458b, as Figure~\ref{wasp6b}. HD
  209458b has the strongest evidence in favour of a grey cloud model.\label{hd209458b}}
\end{figure*}

\begin{figure*}
\centering
\includegraphics[width=0.85\textwidth]{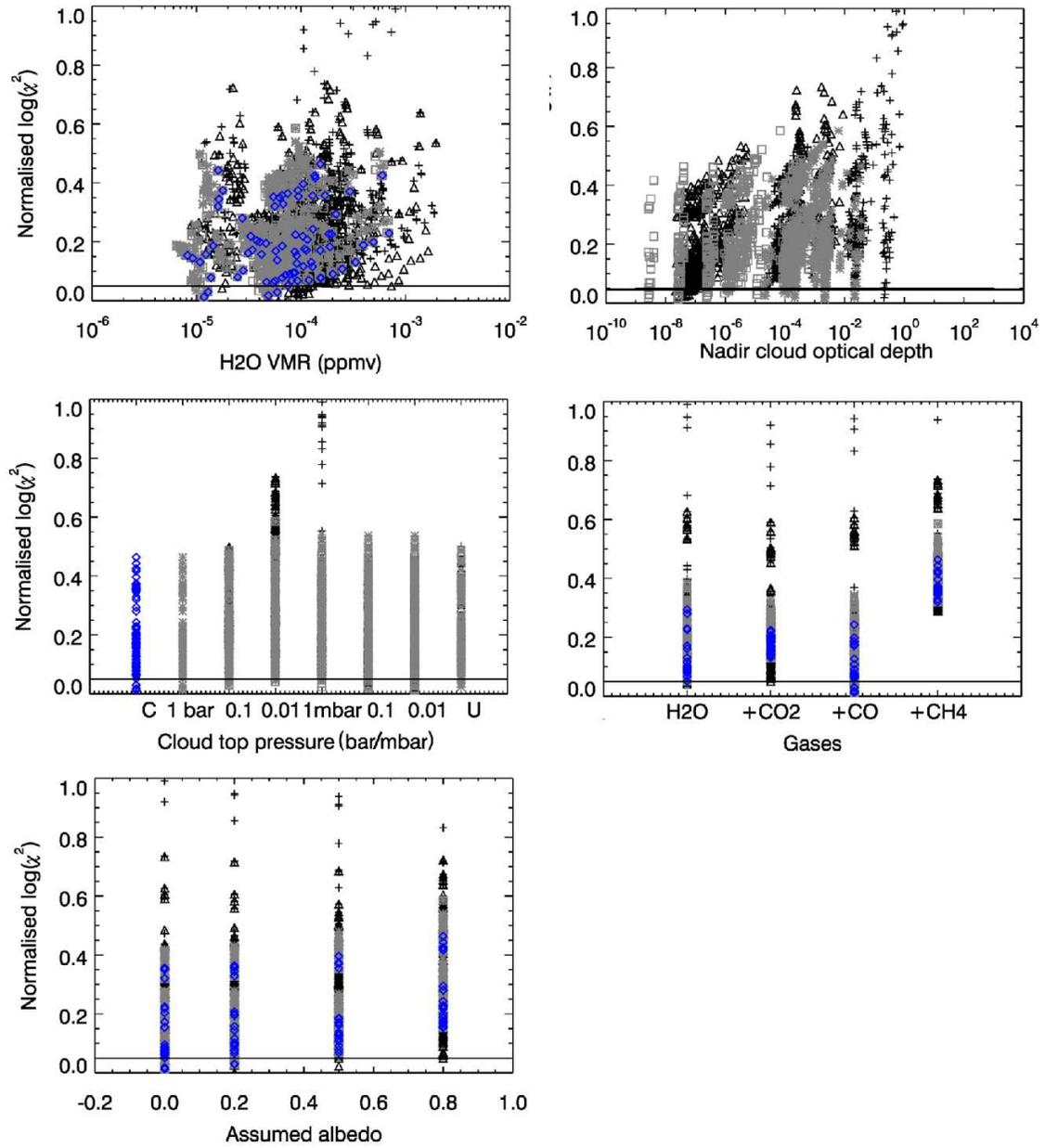}
\caption{Full results for WASP-39b, as Figure~\ref{wasp6b}. The poor
  constraint on H$_2$O abundance is due to the lack of WFC3 data.\label{wasp39b}}
\end{figure*}

\begin{figure*}
\centering
\includegraphics[width=0.85\textwidth]{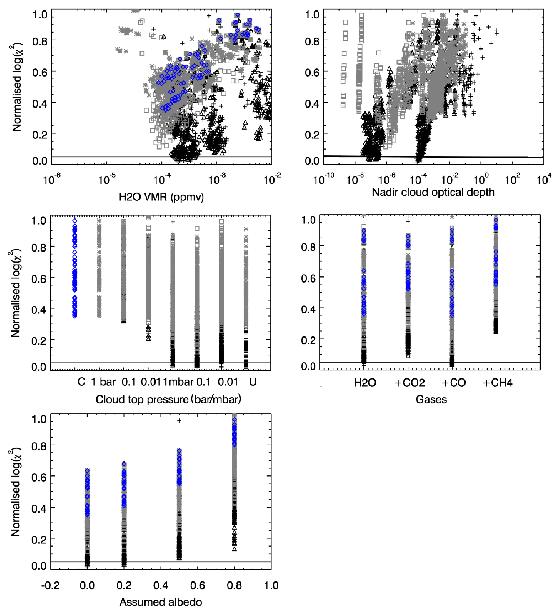}
\caption{Full results for WASP-17b, as Figure~\ref{wasp6b}.\label{wasp17b}}
\end{figure*}

\appendix{SUPPLEMENTARY MATERIAL}
\label{supp}
In this supplementary section we include plots of model parameter
values as a function of reduced $\chi^2$ for each of the 3,600 runs
for each planet (Figures~\ref{wasp6b}---\ref{wasp17b}). These plots indicate the degree to which each
parameter can be determined; broad spreads of values indicate a lack
of constraint. For H$_2$O abundance, three clusters of points
correspond to three different \textit{a priori} values used; where
points are evenly distributed between three clusters, this indicates a
lack of information about H$_2$O abundance in the spectrum. 

Many of the planets have bi-modal solutions for nadir cloud optical
depth. These two solutions correspond to the extended and
decade-confined cloud models for each top pressure, since the nadir
optical depth for a decade-confined model is generally much lower as
the deep regions of the atmosphere are clear.


\label{lastpage} \end{document}